# Lattice thermal expansion and anisotropic displacements in urea, bromomalonic aldehyde, pentachloropyridine and naphthalene

Janine George,[1] Ruimin Wang,[1] Ulli Englert,[1,*] and Richard Dronskowski[1,2,*]

[1] Institute of Inorganic Chemistry, RWTH Aachen University, 52056 Aachen, Germany. E-mail: drons@HAL9000.ac.rwth-aachen.de; ullrich.englert@ac.rwth-aachen.de

[2] Jülich-Aachen Research Alliance (JARA-HPC), RWTH Aachen University, 52056 Aachen, Germany

**Abstract:**

Anisotropic displacement parameters (ADPs) are commonly used in crystallography, chemistry and related fields to describe and quantify thermal motion of atoms. Within the very recent years, these ADPs have become predictable by lattice dynamics in combination with first-principles theory. Here, we study four very different molecular crystals, namely urea, bromomalonic aldehyde, pentachloropyridine, and naphthalene, by first-principles theory to assess the quality of ADPs calculated in the quasi-harmonic approximation. In addition, we predict both thermal expansion and thermal motion within the quasi-harmonic approximation and compare the predictions with experimental data. Very reliable ADPs are calculated within the quasi-harmonic approximation for all four cases up to at least 200 K, and they turn out to be in better agreement with experiment than the harmonic ones. In one particular case, ADPs can even reliably be predicted up to room temperature. Our results also hint at the importance of normal-mode anharmonicity in the calculation of ADPs.

## I. INTRODUCTION

Thermal motion of atoms is a property of widespread interest both in the physical and chemical communities.[1-6] In crystallographic studies, such atomic thermal motion is typically quantified by anisotropic displacement parameters (ADPs) and should be considered to arrive at reliable crystal structures. X-ray diffraction, by far the most popular diffraction technique, determines the distribution of electron density. Even in the absence of systematic errors such as absorption, thermal motion is not the only reason for "smearing out" electron density in real space, and ADP refinement has served to nicely lower the residual values by introducing plenty of additional parameters in the first place. Other unwanted effects such as atomic disorder may somehow pop up from fairly anisotropic thermal ellipsoids, the "ORTEP cigar" or "ORTEP saucer" scenarios – which immediately questions the available structural model. Nonetheless, given a constantly increasing overall quality of X-ray and also neutron diffraction intensities, there is a lot of scientific data to be extracted from properly refined ADPs. At the same time, first-principles quantum theory allows for an independent perspective because calculating ADPs without prior information has become possible.

In recent work, such first-principles ADPs have been computed for a wide range of materials: metal carbides,[7] cryolite,[4] layered chalcogenides,[8] diamond,[9] metal-organic frameworks,[10] organometallic compounds,[11] and molecular crystals.[5, 11-14] Thus far, the ADP calculations have mostly been done by lattice dynamics in the harmonic approximation; first-principles thermal expansion has only been considered in the ADP computation for one prototypical dispersion-dominated compound, namely α-sulfur.[11] For the latter molecular crystal, including thermal expansion by using a quasi-harmonic approximation improved the ADP calculation up to 200 K when a suitable dispersion-corrected DFT method was applied. In addition, lattice-dynamics calculations have enabled new refinement strategies for diffraction data in the recent years.[15, 16]

In this work, we now study four very different molecular crystals in terms of their intermolecular interactions: urea $CH_4N_2O$ **1**,[17-19] bromomalonic aldehyde $BrC_3H_3O_2$ **2**,[20-23] pentachloropyridine $C_5Cl_5N$ **3**,[13, 24] and naphthalene $C_{10}H_8$ **4**.[25] This should provide us with a deeper insight in how thermal expansion influences the ADP calculation for a selected diversity of molecular crystals. Moreover, we check the influence of normal-mode anharmonicity on ADPs of urea and naphthalene calculated at the Γ-point.

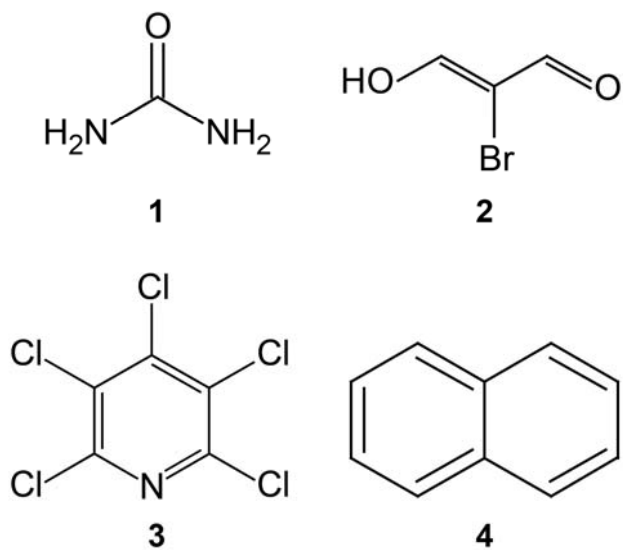

Scheme 1.

To start with, urea crystallizes in the tetragonal space group $P\bar{4}2_1m$ (Figure 1a).[17-19] All atoms occupy special positions, with O and C in Wyckoff position 2c and N and H in 4e, hence each molecule displays crystallographic $C_{2v}$ symmetry. Strong N−H···O hydrogen bonds connect neighboring molecules. Earlier reports on lattice-dynamics calculations in the framework of DFT treated urea's thermal ellipsoids in the harmonic approximation[12, 14] and its thermal expansion in the quasi-harmonic approximation.[26] Nonetheless, no calculations of thermal ellipsoids for urea within the quasi-harmonic approximation have been communicated. Studies on urea's charge density and the experimental analysis of its thermal motion suggest that some atoms do show contributions beyond simple harmonic thermal motion, even at 123 K.[27]

Bromomalonic aldehyde adopts the orthorhombic space group $Cmc2_1$[20-23] (Figure 1b) for which all atoms are located in Wyckoff positions 4a. This implies that the necessarily planar molecules are organized in layers parallel to the (011) plane and connect with each other by resonance-assisted hydrogen bonds to form chains extending in [001] direction. Adjacent chains within the layers interact via weak hydrogen bonds and also halogen bonds along [010].

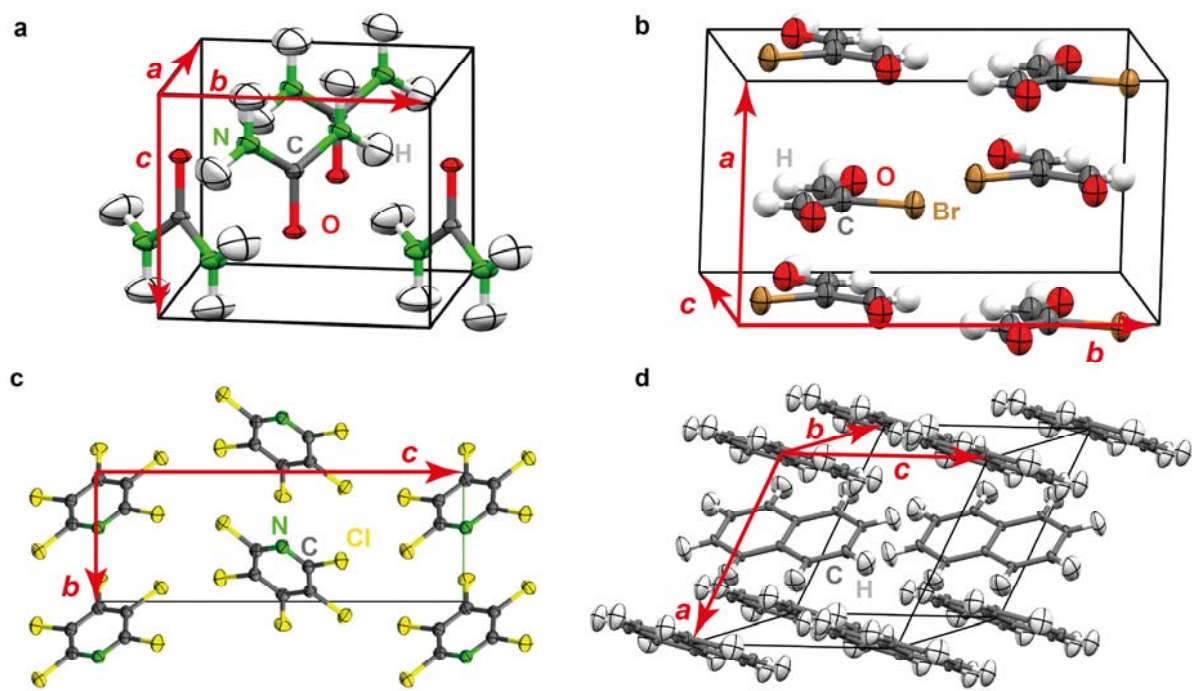

Fig 1. Crystal structures of a) urea at 12 K,[17] b) bromomalonic aldehyde at 100 K,[20] c) pentachloropyridine at 100 K[13] and d) naphthalene at 5 K.[28] The crystal structures were visualized with Mercury [29] and the displacement ellipsoids are shown at the 90% probability level.

We have already conducted a computational analysis of the intermolecular interactions in bromomalonic aldehyde by applying an energetic partitioning scheme, and we have compared these results to single-crystal diffraction experiments at high resolution. In addition, we have been able to correlate the different interaction types with experimental data for the strongly anisotropic thermal expansion of this layered structure.[20]

Pentachloropyridine shows dimorphism.[13, 24] Here we only discuss the monoclinic polymorph because crystals of high quality and thus high-quality ADPs could only be obtained for this very polymorph.[13] It is monoclinic and crystallizes in space group $Pc$ where all atoms sit on general positions (see Figure 1c). Moreover, neighboring molecules form an N···Cl halogen bond – the only intermolecular interaction shorter than the sum of the van-der-Waals radii in this compound. We have already predicted ADPs at several levels of theory within the harmonic approximation for this solid. [13] Their agreement with experiment was sufficient up to 150 K.

Finally, naphthalene adopts the most common space group for molecular crystals, $P2_1/a$ [25] (Fig. 1d) in which the aromatic molecules sit on a crystallographic inversion center. In contrast to the aforementioned three crystals, the naphthalene crystal is mainly held together by van-der-Waals interactions.[30] Its thermal expansion and the thermal motion of its atoms are very well investigated.[1, 25, 30-35]

## II. THEORY

For all phonon calculations and structural optimizations, the forces were calculated using dispersion-corrected DFT as implemented in VASP,[36-38] with strict convergence criteria of $\Delta E < 10^{-7}$ ($10^{-5}$) eV per cell for electronic (structural) optimizations, respectively. Moreover, we used the projector augmented-wave[39, 40] method with a plane wave cutoff of 500 eV and the following functionals with dispersion-corrections: the Perdew–Burke–Ernzerhof (PBE) functional augmented by the "D3" correction of Grimme and co-workers together with Becke–Johnson damping, denoted PBE+D3(BJ) in the following.[41-43] Besides the traditional damping parameters as fitted by Grimme and co-workers ($s_6 = 1.00$, $s_8 = 0.787500$, $\alpha_1 = 0.428900$, $\alpha_2 = 4.440700$), we also used those by Sherrill et al.[44] ($s_6 = 1.00$, $s_8 = 0.358940$, $\alpha_1 = 0.012092$, $\alpha_2 = 5.938951$); we denote this method PBE+D3M(BJ). The latter is more reliable for dispersion-dominated organic complexes[44] and it is therefore expected to be also more reliable for molecular crystals dominated by dispersion interactions.

For the phonon calculations, we calculated harmonic vibrational frequencies and normalized phonon eigenvectors by using the finite displacement method as implemented in Phonopy with a displacement of 0.01 Å.[45, 46] To do so, we used 3×3×4 supercells for urea, 3×2×3 supercells for bromomalonic aldehyde, 4×4×2 supercells for pentachloropyridine and 2×3×2 supercells for naphthalene; all calculations were performed at the Γ-point. Afterwards, vibrational frequencies and eigenvectors were used to compute the vibrational part of the Helmholtz free energy $A_\text{vib}$ and also to calculate the mean-square displacement matrices as implemented in Phonopy.[12, 45, 46] We also checked the influence of the supercell size on the calculated ADPs (see Supplemental Material).

To predict thermal expansion, we applied the quasi-harmonic approximation, as also implemented in Phonopy. [45-47] In doing so, we started with the energetically fully optimized structure at the ground-state volume ($V_0$). Then, we calculated an energy–volume curve $E_0(V)$ at 13 volumes ranging from $0.94^3 \times V_0$ to $1.06^3 \times V_0$, with a constant-volume optimization at each of the volumes. The lattice parameters and atom positions were thereby optimized by minimizing the electronic energy.

The vibrational part of the Helmholtz free energy $A_{\text{vib}}$ was then calculated at the 13 volumes ranging from $0.94^3 \times V_0$ to $1.06^3 \times V_0$ and at several temperatures $T$. Next, the Helmholtz free energy $A(V;T)$ was calculated by

$$A(V;T) = E_0(V) + A_{\text{vib}}(V;T). \qquad (1)$$

Fitting the Vinet equation of state [48] to $A(V;T)$ and minimizing the resulting equation of state $A'(V;T)$ at each temperature gave the optimal volume at the different temperatures:

$$V_0(T) = \arg\min_{V}(A'(V;T)). \qquad (2)$$

To determine lattice parameters corresponding to each of the volumes, we used VASP to relax the structures in order to minimize the energy under the constraint of constant volumes $V_0(T)$, as described in previous work.[11] This results in the evaluation of the following equation for a monoclinic crystal system:

$$(a_0(T), b_0(T), c_0(T), \beta_0(T)) = \arg\min_{(a,b,c,\beta)}(E_0(a,b,c,\beta)), \text{ subject to } abc\sin(\beta) = V_0(T)$$

(3)

$E_0(a, b, c, \beta)$ represents the DFT energy. This implies that the derived lattice parameters minimize the DFT energy and not the free energy for a given volume $V_0(T)$. Such approximations to describe the evolution of the lattice parameters are very common in the quasi-harmonic approximation applied to crystals with non-cubic crystal systems; they result in good agreement with thermodynamic data and are thus justified.[26, 49-51] In contrast, a correct minimization of $A(a, b, c, \beta)$ is computationally extremely demanding.

Next, we calculated mean-square displacement matrices at the expanded volumes and ground-state volumes. During this calculation frequencies lower than 0.10 THz were cut off for urea, naphthalene and bromomalonic aldehyde and lower than 0.13 THz for pentachloropyridine such as to reduce computational noise as described in previous work.[13] The mean-square displacement matrices were afterwards transformed into anisotropic displacement tensors and the Cartesian coordinate system was changed back to the crystal coordinate system.[13, 52]

Also, we computed their main-axis components $U_1$, $U_2$, and $U_3$ which were used to calculate the equivalent isotropic displacement parameter $U_{\text{eq}}$, a value comparable to the isotropic displacement parameter $U_{\text{iso}}$:

$$U_{\text{eq}} = \frac{U_1 + U_2 + U_3}{3}. \qquad (4)$$

This was conducted by a custom-written program [13, 53] based on the formulas in Ref. 52.

To check for further anharmonicity, we used a frozen-phonon *ansatz* for urea and naphthalene at the Γ-point, equivalent to a non-interacting phonons scenario. We displaced the atoms along each of the phonon modes at the Γ-point, calculated an energy-displacement curve and fitted the latter by a polynomial of up to 28$^{th}$ order ($E=a_2 \times x^2 + a_3 \times x^3 + a_4 \times x^4 + [\ldots] + a_{28} \times x^{28}$) to eventually solve the resulting 1D Schrödinger equation. The displaced structures were generated with PHONOPY by using the MODULATION tag, and we converged the anharmonic frequencies by increasing the order of the polynomial. A comparison of the harmonic frequencies derived from the quadratic part of this fit to the result from the finite displacement method is given in the supplement. The frozen-phonon *ansatz* and the derivation of the effective mass were done accordingly to ref [54]: at an absolute total displacement $u_{\text{ges},j} = \sum_k \sqrt{\boldsymbol{u}_{k,j}^2} = 1$ along a phonon mode with $u_{k,j}$ as a displacement of an atom $k$, the effective mass of the phonon mode $j$ is defined as follows:

$$m_{\text{eff},j} = \sum_k m_k \boldsymbol{u}_{k,j}^2. \quad (5)$$

$u_{\text{ges},j}$ is connected to the normal-mode coordinate $q_j$ as described in the following equation:

$$q_j = \sqrt{m_{\text{eff},j}} u_{\text{ges},j}. \quad (6)$$

The 1D-Schrödinger equation is solved by using the procedure described in 55, a method based on representing the momentum and the position operator in a discrete matrix by utilizing the normalized eigenstates of a harmonic oscillator. Thus, the size of these discrete matrices must be converged. The ADPs based on this frozen-phonon method were then calculated as described in ref. 4 (equations 18 to 22 therein). Such frozen-phonon approaches would also enable us to treat soft modes in the ADP calculation.[4, 56] Further details on this frozen-phonon approach can be found in the supplement.

## III. EXPERIMENT

Single crystals of bromomalonic aldehyde were obtained by slow isothermal evaporation from acetonitrile. A specimen with approximate dimensions 0.30 × 0.16 × 0.16 mm was placed in the cold N$_2$ stream of an Oxford Cryostream 700 device. Data were collected on a Stoe & Cie Stadivari goniometer equipped with a DECTRIS Pilatus 300k detector. Intensities were integrated and corrected for absorption with the X-Area[57] suite. The known crystal structure of bromomalonic aldehyde[20-22] served as input for a least-squares refinement on $F^2$.[58] The quality of the intensity data on which our experimental ADPs are based outperforms our previously reported[20] data set with respect to resolution, redundancy and agreement factors. Key quality criteria: resolution $\sin\theta_{max}/\lambda$ = 1.18 Å$^{-1}$, 17877 reflections, 3025 independent intensities, $R_{int}$ = 0.0185, $R$ = 0.0256, $wR_2$ = 0.0490 for all data, ratio of reflections/parameters > 80, maximum fluctuations in a final Fourier difference synthesis 0.49/−1.27 eÅ$^{-3}$, ratio between maximum and minimum principal components in the most anisotropic displacement parameter 2.50. The results have been deposited in CIF format, CCDC 1548810.

## IV. RESULTS AND DISCUSSION

### A. Ground state

First, let us evaluate the ground states derived; for this purpose, the structures were optimized minimizing the total energy. For urea and naphthalene, both the calculations at the PBE+D3(BJ) and PBE+D3M(BJ) arrive at lattice parameters and volumes in very good agreement with experiment (Table 1). As expected, the volumes of the ground states are smaller than the experimental ones at very low finite temperature (< 15 K). This is different for bromomalonic aldehyde and pentachloropyridine. Both ground-state volumes at the PBE+D3(BJ) level overestimate the experimental ones at 20 and 100 K, respectively (Table 1). In contrast, the volumes at the PBE+D3M(BJ) level are in better agreement with experiment. These tendencies are also mostly reflected by the root mean square of the Cartesian displacements calculated according to ref. 59.

Fortunately, data from vibrational spectroscopy for urea and from inelastic neutron diffraction for naphthalene at low temperature are available.[60, 61] Thus, we may compare the vibrational frequencies calculated at the ground-state volume and in the harmonic approximation to the experimental data (Fig. 2 and Fig. 3). Both methods are in very good agreement with experiment but PBE+D3M(BJ) overestimates the frequencies of naphthalene slightly more than PBE+D3(BJ). However, we note that we only compare the ground states here.

For bromomalonic aldehyde and monoclinic pentachloropyridine, no vibrational data are available up to now. However, we include the vibrational frequencies and their irreducible represen-tation at the Γ-point for all four compounds in the Supplementary Material. Moreover, we analyze the contributions of each atom to each vibrational state at the Γ-point closely so that future vibrational spectroscopy data may be understood more easily.[62, 63]

Table 1. Lattice parameters and volume per formula unit of crystalline urea, bromomalonic aldehyde, monoclinic pentachloropyridine, and naphthalene obtained by optimization of the total energy at different levels of theory (not including ZPE corrections) and from experiment (urea:[17] single-crystal neutron diffraction at 12 K, bromomalonic aldehyde:[20] powder X-ray diffraction at 20 K, pentachloropyridine:[13] single-crystal X-ray diffraction at 100 K, naphthalene:[25] single-crystal neutron diffraction at 5 K), as well as the root mean square (RMS) of the Cartesian deviations between calculated structures and the experimental ones.[59, 64] For X-ray diffraction data, we did not include the hydrogen atoms in the RMS calculation.

| **Urea** | $a$ (Å) | $b$ (Å) | $c$ (Å) | $\beta$ (°) | $V/Z$ (Å$^3$) | RMS (Å) |
|---|---|---|---|---|---|---|
| Expt. (12 K)[17] | 5.565(1) | =$a$ | 4.684(1) | 90 | 72.53 | − |
| PBE+D3M(BJ) | 5.485 | =$a$ | 4.664 | 90 | 70.16 | 0.03 |
| PBE+D3(BJ) | 5.487 | =$a$ | 4.665 | 90 | 70.22 | 0.03 |
| **Bromomalonic Aldehyde** | $a$ (Å) | $b$ (Å) | $c$ (Å) | $\beta$ (°) | $V/Z$ (Å$^3$) | RMS (Å) |
| Expt. (20 K)[20] | 6.288(2) | 10.730(4) | 6.399(2) | 90 | 107.9 | − |
| PBE+D3M(BJ) | 6.235 | 10.596 | 6.368 | 90 | 105.2 | 0.04* |
| PBE+D3(BJ) | 6.352 | 10.696 | 6.382 | 90 | 108.4 | 0.04* |
| **Pentachloro-pyridine** | $a$ (Å) | $b$ (Å) | $c$ (Å) | $\beta$ (°) | $V/Z$ (Å$^3$) | RMS (Å) |
| Expt. (100 K)[13] | 5.3122(2) | 5.1770(2) | 14.8307(6) | 99.493(2) | 201.14(2) | − |
| PBE+D3M(BJ) | 5.263 | 5.136 | 14.824 | 99.58 | 197.57 | 0.02 |
| PBE+D3(BJ)[13] | 5.327 | 5.176 | 15.188 | 100.27 | 206.45 | 0.05 |
| **Naphthalene** | $a$ (Å) | $b$ (Å) | $c$ (Å) | $\beta$ (°) | $V/Z$ (Å$^3$) | RMS (Å) |
| Expt. (5 K)[25] | 8.080(5) | 5.933(2) | 8.632(5) | 124.65(4) | 170.21 | − |
| PBE+D3M(BJ) | 7.955 | 5.868 | 8.585 | 124.6 | 164.94 | 0.03 |
| PBE+D3(BJ) | 7.996 | 5.889 | 8.585 | 124.4 | 166.77 | 0.02 |

*RMS value is relative to the experimental 100 K structure (this work).

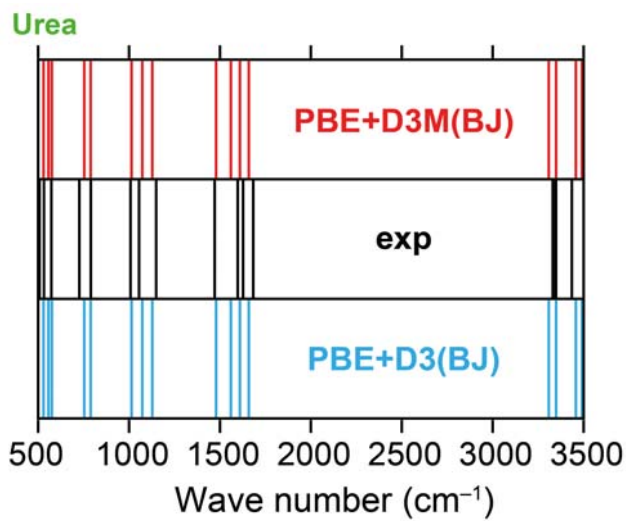

Fig. 2. Comparison of all IR-active modes (irreducible representations: *B2* and *E*) from theory and experiment at 77 K.[60] The results at the PBE+D3M(BJ) and PBE+D3(BJ) level are nearly indistinguishable. IR-active irreducible representations were derived by the program SAM [65] on the Bilbao Crystallographic Server.

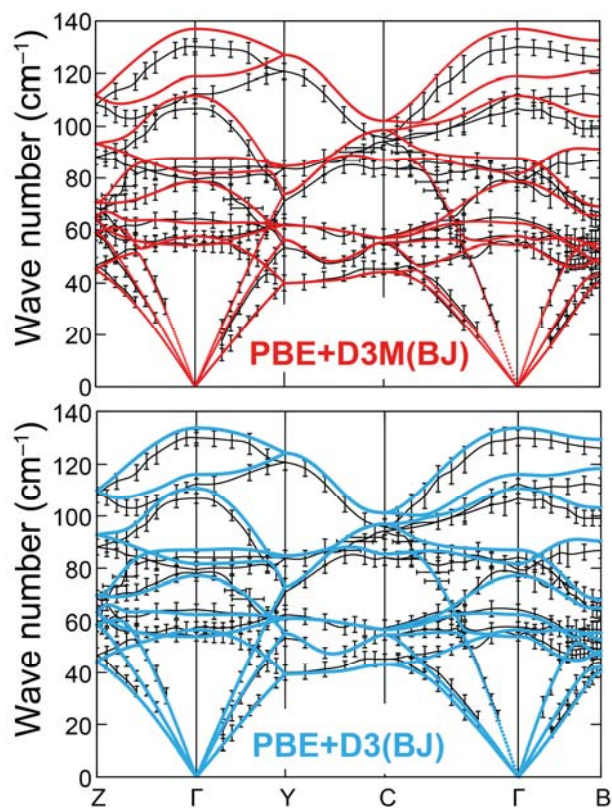

Fig 3. Fully deuterated naphthalene's phonon band structure from the PBE+D3(BJ) and PBE+D3(BJ)M level of theory in the harmonic approximation compared to the experimentally derived one by inelastic neutron scattering at 6 K.[61] Error bars correspond to three standard uncertainties. The graph is based on the same force calculation as the ADP calculations later on, but the mass is changed from +1.008 (H) to

+2.014 (D). The PBE+D3(BJ)-based phonon band structure is in slightly better agreement with the experimental data. The reciprocal space was analyzed with the help of the Brillouin-zone database on the Bilbao Crystallographic Server.[66] The ground-state volume of PBE+D3M(BJ) is roughly 1% smaller than that of PBE+D3(BJ). The zero-point energy has not been considered.

## B. Thermal Expansion

We now turn to temperature-dependent volumes from the quasi-harmonic approximation and compare them to experimentally derived ones (Fig. 4).

As before, let us start with urea. Both DFT methods provide nearly the same volumes and overestimate the volume $V$ slightly ($\Delta V < 2$ % at 150 K; Figure 4). As a side note, the densities of phonon states show small contributions of imaginary modes (up to 1.4% of the all vibrational states), in particular for volumes smaller than the equilibrium volume. Urea is known for several high-pressure polymorphs;[67] this finding might not only allude to these high-pressure structural alternatives, it might also show up by a slightly worse quality of the quasi-harmonic approximation for urea. However, this influence of small traces of soft modes is usually very low.[68]

For naphthalene, both methods also lead to slightly larger volumes than the experimental ones, especially at higher temperatures. PBE+D3M(BJ) generates volumes in slightly better agreement with experiment.

The performance of both DFT methods differs strongly for both bromomalonic aldehyde and pentachloropyridine. Here, PBE+D3M(BJ) is clearly superior to PBE+D3(BJ) (bromomalonic aldehyde: $\Delta V(250$ K$) = 2.2$% vs. 5.5%; pentachloropyridine: $\Delta V(300$ K$) = 2.9$% vs. 7.9% ). As before, the volumes at higher temperatures are in less satisfactory agreement with experiment.

All slopes of the volume-temperature curves are significantly overestimated (Table 2). For pentachloropyridine, naphthalene and bromomalonic aldehyde, the relative overestimation of the experimental slopes is slightly lower (30−40%) than for urea (65%). Moreover, the absolute overestimation is most pronounced for naphthalene. Again, PBE+D3M(BJ) is clearly superior for pentachloropyridine and bromomalonic aldehyde. This might be explainable by the larger and also different training that was used to fit the damping parameters of the D3M(BJ) method in comparison to the D3(BJ) method. [42-44] Note that the D3M(BJ) training set includes many more data points originating from organic halides and halogen complexes than the D3(BJ) training set.

In the case of bromomalonic aldehyde, we also look at the anisotropic thermal expansion (Figure 5 and Table 3) in order to investigate direction- and bond-character dependence. Both methods are able to reproduce the qualitative differences in the thermal expansion of the lattice parameters. Along *a*, the molecules are connected by electrostatic and dispersion interactions and, therefore, the lattice expansion is largest. Along *b*, the molecules are connected by weak hydrogen bonds and halogen bonds, so that thermal expansion is smaller. Unfortunately, this strong difference between the *a* and *b* directions is not very well described by theory; the theoretical slopes of the *a*- and *b*-resolved temperature curves are very similar, in strong contrast to the experimental ones. Along *c*, resonance-assisted hydrogen bonds connect the molecules, so the thermal expansion along *c* is therefore nearly zero; accordingly, the theoretical value is also lower but still overestimates the effect. As for the volume expansion, PBE+D3M(BJ) shows better agreement with the experimental anisotropic thermal expansion than PBE+D3(BJ). Moreover, both methods are at least in qualitative agreement with experiment.

Summing up, the thermal expansion is overestimated by the quasi-harmonic approximation for all systems. We have seen this for α-sulfur as well and assume that this effect originates partly from additional anharmonicity.[11, 69, 70] For all further evaluations, we continue at the PBE+D3M(BJ) level because it agrees better with experiment for pentachloropyridine and bromomalonic aldehyde.

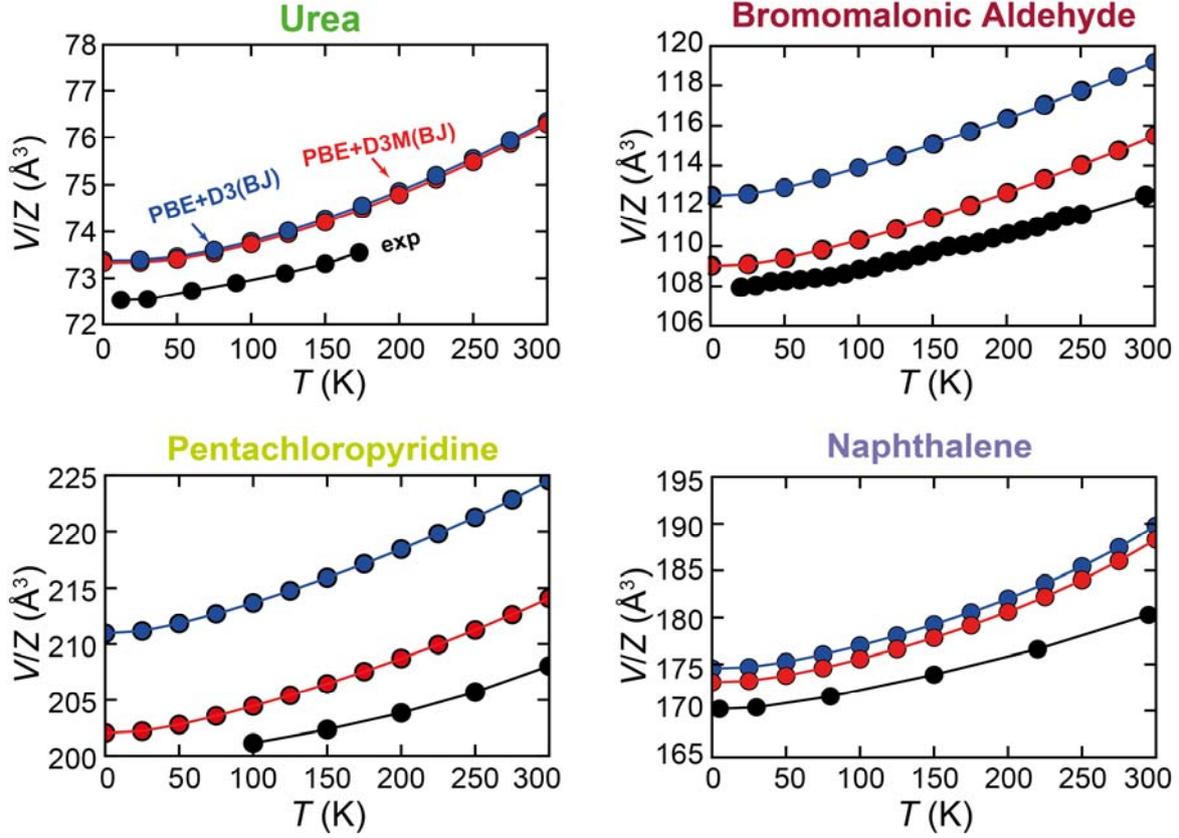

Figure 4. Thermal expansion calculated in the quasi-harmonic approximation compared to experimental data for urea,[17] bromomalonic aldehyde,[20] pentachloropyridine,[13] and naphthalene.[25]

Table 2. Slope of the volume-temperature plot $\Delta V/\Delta T$ of urea,[17] bromomalonic aldehyde,[20] pentachloropyridine,[13] and naphthalene[25] derived by linear regression of the volume in the temperature-range from 50 to 300 K.

| $\Delta V/\Delta T$ (Å³/K) | Expt. | PBE+D3(BJ) | PBE+D3M(BJ) |
|---|---|---|---|
| Urea | $(0.72\pm0.05)\times10^{-2}$ | $(1.17\pm0.05)\times10^{-2}$ | $(1.16\pm0.05)\times10^{-2}$ |
| Bromomalonic aldehyde | $(1.78\pm0.04)\times10^{-2}$ | $(2.52\pm0.05)\times10^{-2}$ | $(2.46\pm0.06)\times10^{-2}$ |
| Pentachloropyridine | $(3.43\pm0.26)\times10^{-2}$ | $(5.07\pm0.16)\times10^{-2}$ | $(4.53\pm0.13)\times10^{-2}$ |
| Naphthalene | $(4.06\pm0.26)\times10^{-2}$ | $(5.73\pm0.29)\times10^{-2}$ | $(5.74\pm0.28)\times10^{-2}$ |

Table 3. Slopes of lattice parameter vs temperature plots of bromomalonic aldehyde, derived by linear regression of the lattice parameters $a$, $b$, and $c$ in the temperature-range from 50 to 300 K.

| | Expt. | PBE + D3(BJ) | PBE + D3M(BJ) |
|---|---|---|---|
| $\Delta a/\Delta T$ (Å/K) | $(8.24\pm0.15)\times10^{-4}$ | $(8.95\pm0.18)\times10^{-4}$ | $(8.96\pm0.26)\times10^{-4}$ |
| $\Delta b/\Delta T$ (Å/K) | $(3.77\pm0.12)\times10^{-4}$ | $(8.09\pm0.20)\times10^{-4}$ | $(7.80\pm0.14)\times10^{-4}$ |
| $\Delta c/\Delta T$ (Å/K) | $(-0.134\pm0.050)\times10^{-4}$ | $(0.552\pm0.023)\times10^{-4}$ | $(0.555\pm0.017)\times10^{-4}$ |

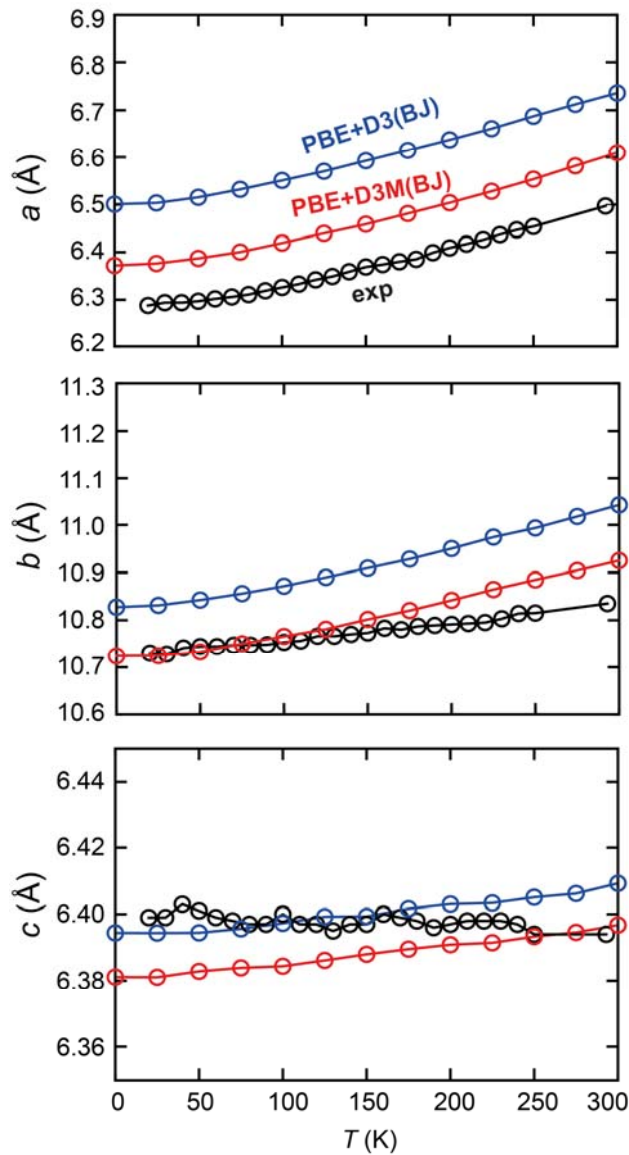

Figure 5. Anisotropic lattice expansion for bromomalonic aldehyde.[20] Both DFT methods are able to reproduce the qualitative anisotropic lattice expansion for bromomalonic aldehyde. Note the enlarged scale for the $c$ axis in the bottom panel.

### C. ADPs at expanded volumes

Next, let us calculate ADPs in both the harmonic and quasi-harmonic approximation for all four systems.

We start by assessing the ADPs calculated for urea at 12, 60 and 123 K (Figure 6). At 12 and 60 K, the results from the quasi-harmonic approximation are in slightly better agreement with experiment than those from the harmonic approximation. At 123 K, this tendency changes; the quasi-harmonic approximation overestimates the experimental ADPs slightly. This originates mainly from the overestimation of the ADP of the nitrogen atoms both in the harmonic and the quasi-harmonic approximation, and it may go back to the anharmonicity mainly originating from the hydrogen bonds. We will come back to this assumption by performing frozen-phonon calculations including anharmonicity at the Γ-point, see below.

For bromomalonic aldehyde and pentachloropyridine, the quasi-harmonic ADPs are clearly in better agreement with experiment than the harmonic ones. In some cases, they even fit perfectly to the experimental values. For the two cases of bromomalonic aldehyde and pentachloropyridine, there is an obvious correlation between the good agreement with the experimental volume expansion and good ADPs.

For naphthalene, the entire situation looks quite different: the harmonic approximation arrives at ADPs in better agreement with experiment for nearly all investigated temperatures (5, 80, 150, 220 and 295 K); the tendency is only changed for 30 K. For brevity, we only present the ADPs calculated at 5, 80 und 295 K. The ADPs at 5 and 80 K calculated in the quasi-harmonic approximation are at least in sufficient agreement with experiment. The slope of a linear fit between experiment and theory without consideration of an interecept arrives at values between 1.0 and 1.11; the ADPs are thus only slightly overestimated. At 295 K, both harmonic and quasi-harmonic approximation do not fit well to the experimental results; while the harmonic approximation drastically underestimates the experimental values, the quasi-harmonic one drastically overestimates them. The latter finding correlates with a drastic absolute overestimation of the volume expansion. Thus, both levels of theory are in insufficient agreement with experiment at higher temperatures. As side note, the agreement of the H-ADPs in the quasi-harmonic approximation with experiment is slightly better than that of the C-ADPs (slope of linear fit without intercept between experiment and theory at all 6 temperatures: 1.27 (H) vs. 1.32 (C) ). This correlates with the fact that the C-ADPs depend more strongly on low phonon frequencies than the H-ADPs do. Fortunately, especially the estimated H-ADPs are very much needed to complement X-ray diffraction because they cannot be determined by X-ray diffraction and the independent atom model.[71] For organometallic compounds, we have already seen that ligand atoms can be determined with higher confidence than the heavier metal atoms by theory.[11]

As demonstrated by Capelli et al.,[28] ADPs from neutron diffraction include anharmonicity so that the neglect of anharmonicity could play a role in the case of naphthalene. We will test this influence below. In addition, the level of theory could also have deficiencies in describing geometries far away from the equilibrium structure,[44] while the dispersion correction does not include many-body effects which influence naphthalene's lattice energy.[72]

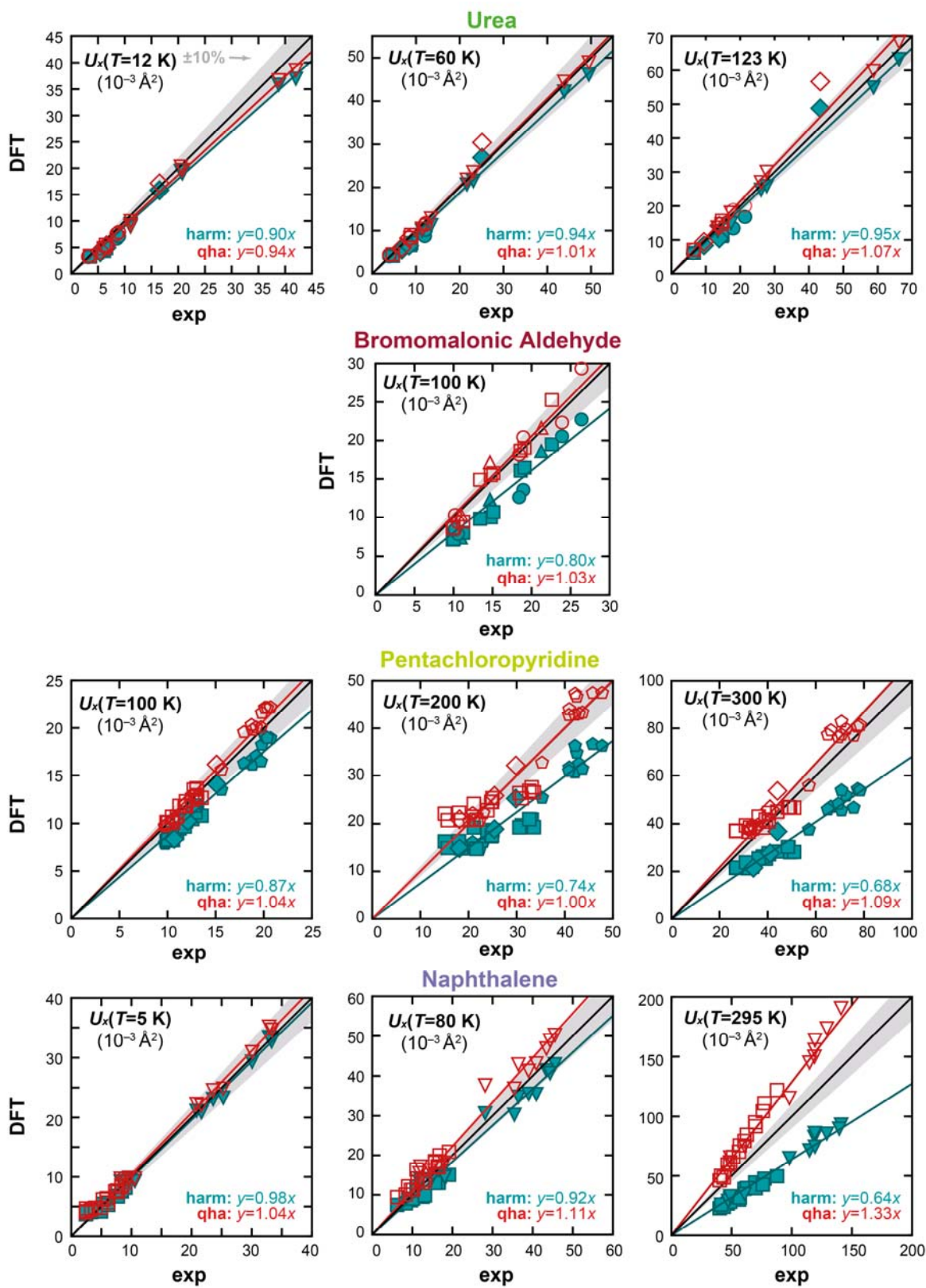

Figure 6. ADPs calculated in the harmonic (harm) and quasi-harmonic approximation (qha) compared to experimental ADPs for urea from neutron diffraction,[17] for bromomalonic aldehyde from X-ray diffraction (this work), for pentachloropyridine from X-ray diffraction[13] and for naphthalene from neutron

diffraction.[25] In the case of bromomalonic aldehyde, no H-ADPs are compared. The atom types are symbolized by ◇ for N, ▽ for H, ◯ for O, □ for C, △ for Br and ⌂ for Cl.

### D. Normal-Mode Anharmonicity and Anisotropic Displacement Parameters

We now turn to the normal-mode anharmonicity of urea and naphthalene and its effect on the calculated ADPs. To save computing time, we only calculated ADPs based on the frequencies at the Γ-point.

Urea will once again serve as the starting case. As depicted in Fig. 7, the quasi-harmonic approximation overestimates the largest main-axis components of N and H more strongly than all other main-axis components. We assess the normal-mode anharmonicity at the 123 K-volume evaluated with the quasi-harmonic approximation.

The most important phonon modes for calculating urea's ADPs at the Γ-point are the two lowest optical phonon modes; they are responsible for nearly 66% of the ADP of N at Γ. As already seen in the evaluation of experimental neutron diffraction data, anharmonicity should clearly influence the size of the ADPs of N and H.[17] The calculated ADPs including and excluding mode anharmonicity of urea at the Γ-point are compared in Figure 7. As expected, ADPs of N and H become smaller if anharmonicity is included; the ADPs of C and O are nearly not influenced. Mostly the anharmonicity of the lowest optical modes are responsible for the smaller ADPs. As said before, for N, the two lowest optical modes are very important; thus, small changes of these oscillators due to anharmonicity can have a huge impact on the overall ADP. For H, the lowest optical modes are still important but amount to only 30% and 15% of the overall ADP in the quasi-harmonic approximation. In Table 4, we first compare the (quasi-)harmonic frequencies of the two lowest optical phonon modes from the finite-displacement method and the frozen-phonon method; they are in very good agreement. Moreover, the fundamental anharmonic frequencies are also given; they grow quite drastically in comparison to the harmonic ones such that the main axes of the ADPs become smaller. The frequencies of the N−H stretch vibrations are also given in Table 5, the frequencies of all other modes are given in the Supplement. We may thus state that the overestimation of urea's ADPs by the quasi-harmonic approximation at 123 K can be partly attributed to the neglect of mode anharmonicity of the lowest phonon modes. In contrast, the slightly better agreement of the harmonic approximation originates from error compensation (neglect of volume expansion and neglect of further anharmonicity).

For napthalene, the influence of the normal-mode anharmonicity on the ADPs at the 150 K-volume evaluated with the quasi-harmonic approximation is depicted in Figure 7. Even at 150 K, the quasi-harmonic approximation overestimates the ADPs. Considering normal-mode anharmonicity at the $\Gamma$-point, however, lowers the size of the ADPs. Therefore, the overestimation of the ADPs by the quasi-harmonic approximation can also be partly attributed to the neglect of this anharmonicity. In contrast to urea, nearly all atoms are affected by the normal-mode anharmonicity.

Mode anharmonicity should also play a role in thermal expansion. Unfortunately, one would have to calculate the anharmonic free energy at several volumes to estimate the influence on the thermal expansion. As one can already guess from the poor quality of ADPs only calculated at the $\Gamma$-point, the anharmonic free energy calculated at the $\Gamma$-point would also be insufficient to calculate the thermal expansion.

In sum, performing such frozen-phonon computations at the $\Gamma$-point enables one to assess the importance of normal-mode anharmonicity for ADP computations of specific compounds. To arrive at anharmonic ADPs in good agreement with experiment, however, one would have to perform the frozen-phonon calculation with supercells to sample more points in the reciprocal space. One might be able to speed up these calculations by only treating the most relevant part, namely the anharmonicity of lower lying phonon modes because their contributions dominate the ADPs.

Table 4. Comparison of the (quasi-)harmonic frequencies of the two lowest optical phonon modes derived by the finite displacement method as implemented in Phonopy ($\omega_{harmonic}$), their (quasi-)harmonic and anharmonic frequencies derived from a displacement of the atoms along each phonon mode ($\omega_{harmonic}$* and $\omega_{anharmonic}$). Moreover, we give deviation between $\omega_{anharmonic}$ and $\omega_{harmonic}$* ($\Delta\omega$) and the Mulliken symbol of the irreducible representations to simplify further comparisons.

| $\omega_{harmonic}$ (cm$^{-1}$) (finite displacement method) | $\omega_{harmonic}$*(cm$^{-1}$) (from frozen-phonon calculation) | $\omega_{anharmonic}$ (cm$^{-1}$) (from frozen-phonon calculation) | $\Delta\omega$ (cm$^{-1}$) | Irreducible representation |
|---|---|---|---|---|
| 55 | 57 | 77 | +20 | *B*1 |
| 68 | 70 | 86 | +16 | *A*2 |

Table 5. As before but for the N−H stretch frequencies.

| $\omega_{harmonic}$ (cm$^{-1}$) (finite displacement method) | $\omega_{harmonic}$*(cm$^{-1}$) (from frozen-phonon calculation) | $\omega_{anharmonic}$ (cm$^{-1}$) (from frozen-phonon calculation) | $\Delta\omega$ (cm$^{-1}$) | Irreducible representation |
|---|---|---|---|---|
| 3337 | 3338 | 3381 | +43 | *E* |
| 3364 | 3363 | 3343 | −20 | *A*1 |
| 3365 | 3364 | 3386 | +22 | *B*2 |
| 3468 | 3467 | 3517 | +50 | *E* |
| 3469 | 3469 | 3488 | +19 | *A*1 |
| 3502 | 3502 | 3525 | +23 | *B*2 |

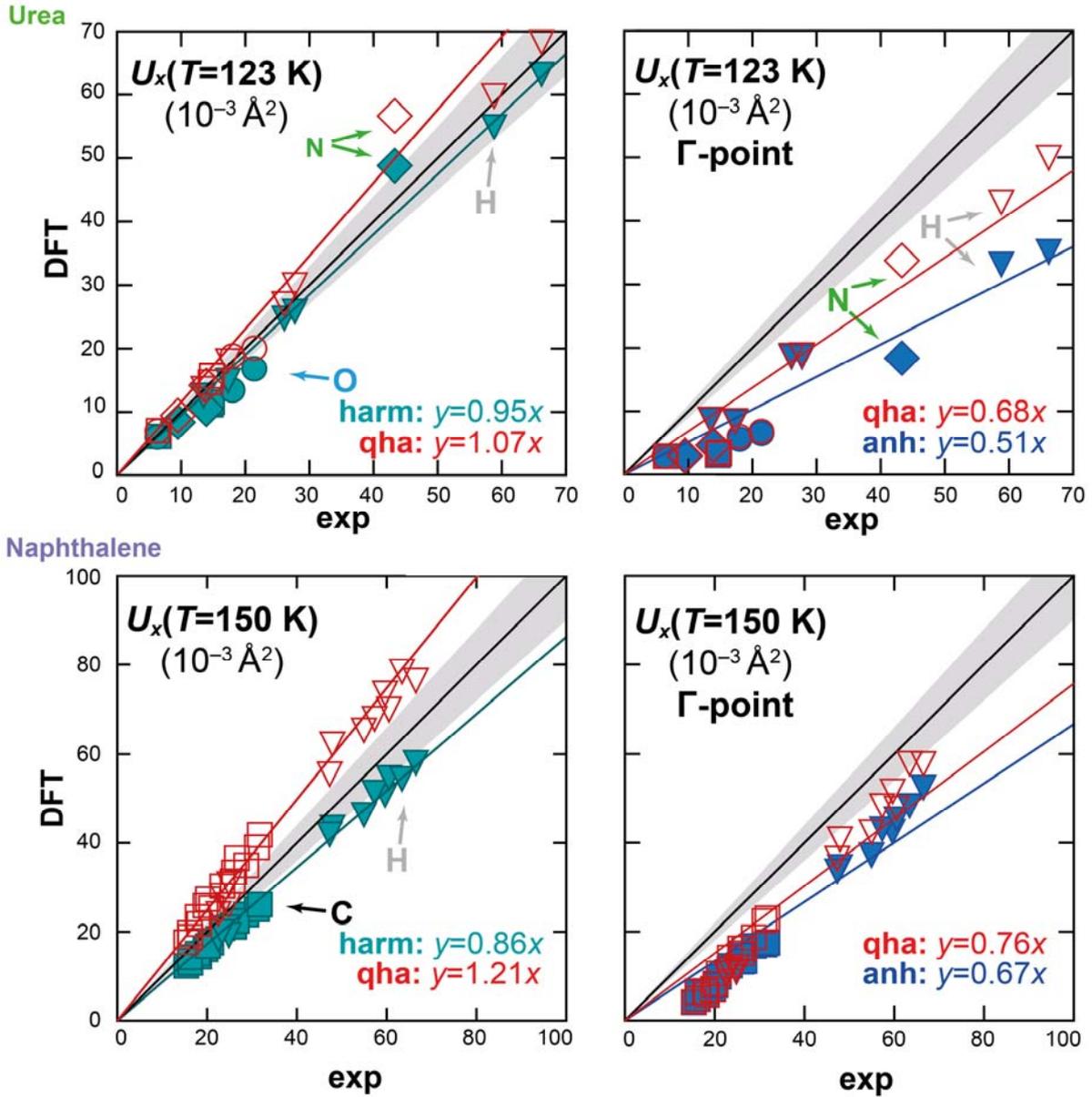

Fig. 7. (Top, left) Comparison of the ADPs of urea at 123 K calculated in the harmonic (harm) and the quasi-harmonic approximation (qha) to experimental data.[17] (Top, right) Comparison of the ADPs of urea at 123 K calculated in the quasi-harmonic approximation (qha) and the ones calculated with a frozen-phonon method (anh) to experimental data.[17] Both calculations are performed at the expanded volume determined by the quasi-harmonic approximation and only at the Γ-point. The latter explains the huge deviations from experiment. (Bottom, left) As above, for naphthalene at 150 K.[25] (Bottom, right) As above for naphthalene at 150 K. In contrast to urea, we have already considered 105 instead of 45 phonon modes. This might partly explain the smaller discrepancies between calculation at the Γ-point and experiment than in the case of urea and the less drastic change of the ADPs due to normal-mode anharmonicity.

## V. CONCLUSIONS

Quantitatively predicting thermal expansion of molecular crystals up to room temperature is still a challenge. For the systems investigated here, the PBE+D3M(BJ) level of theory is in slightly better agreement with experiment than the PBE+D3(BJ) level. The predicted thermal expansion of urea drastically overestimates the experimental one (65%); the ones for naphthalene, pentachloropyridine and bromomalonic aldehyde are in slightly better agreement with experiment (30−40%). As we have shown for bromomalonic aldehyde, strong anisotropic thermal expansion can be at least qualitatively reproduced at this level of theory.

The strong overestimation of the thermal expansion should naturally influence the resulting ADPs from the quasi-harmonic approximation. For naphthalene, the calculated ADPs drastically overestimate the experimental ones above 150 K. In contrast, the results for pentachloropyridine look almost perfect up to room temperature; the absolute overestimation of the thermal expansion is slightly lower than for naphthalene and error compensation might be more fortunate in the latter case.

For urea and bromomalonic aldehyde the ADPs from the quasi-harmonic approximation are in very good agreement with experiment up to 150 K; no reliable experimental ADPs for these compounds at higher temperatures have been reported. In the case of urea and naphthalene, including normal-mode anharmonicity can further improve the agreement with experiment.

Whether reliable ADPs up to room temperature can be calculated obviously depends on the system. Up to 200 K, which we consider a relevant temperature range for the investigation of a variety of molecular crystals, see Figure 8, predicted ADPs in the quasi-harmonic approximation for very different molecular crystals have shown to be consistently good. For this low-temperature range, the less demanding harmonic approximation yields slightly inferior but still satisfactory results. In order to address effects at room temperature, both the DFT level of theory and the description of anharmonicity should be improved.

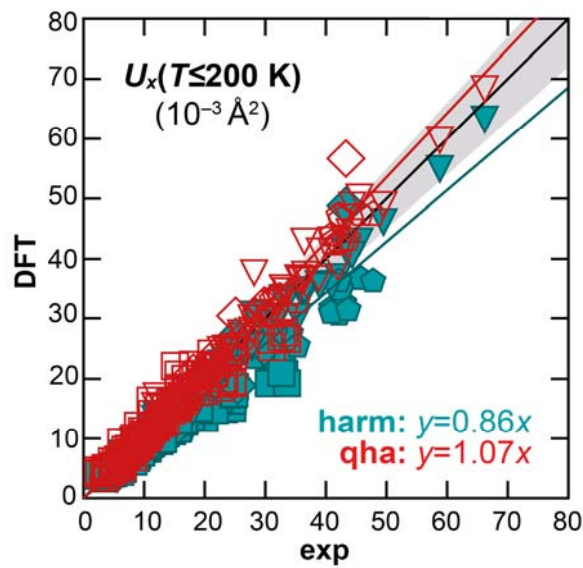

Fig. 8. Comparison of the calculated main-axis components of the anisotropic displacement matrix in the harmonic (harm) and quasi-harmonic approximation (qha) at the PBE+D3M(BJ) level of theory to the experimental values. The plot includes all ADPs for urea at 12, 60 and 123 K[17], for naphthalene at 5, 30, 80 and 150 K[25], for bromomalonic aldehyde at 100 K and for pentachloropyridine at 100, 150 and 200 K[13]. Up to this temperature, even the harmonic approximation seems to be reasonable. We have already demonstrated that the ADP calculation in this temperature regime is reliable for pentachloropyridine at several levels of theory.[13] The atom types are symbolized by ◇ for N, ▽ for H, ○ for O, □ for C, △ for Br and ⬠ for Cl.

## VI. SUPPLEMENTARY MATERIAL

Experimental data on the single-crystal diffraction of bromomalonic aldehyde, additional theoretical details on the density-functional, phonon and ADP calculations.

## VII. ACKNOWLEDGEMENTS

The authors thank Irmgard Kalf for help with crystallization and Dr. Jens Meyer, Stoe & Cie, Darmstadt for collecting diffraction data for bromomalonic aldehyde at high-resolution. Financial support by Fonds der Chemischen Industrie (fellowship for J. G.) and the Deutsche Forschungsgemeinschaft (DFG) is gratefully acknowledged. Computer time was provided by the Jülich–Aachen Research Alliance (Project jara0069).

Supplementary Information for the Manuscript

# Lattice thermal expansion and anisotropic displacements in urea, bromomalonic aldehyde, pentachloropyridine and naphthalene

Janine George,[1] Ruimin Wang,[1] Ulli Englert,[1,]* and Richard Dronskowski[1,2,]*

[1] Institute of Inorganic Chemistry, RWTH Aachen University, Landoltweg.1, Aachen 52074, Germany. E-mail: drons@HAL9000.ac.rwth-aachen.de; ullrich.englert@ac.rwth-aachen.de

[2] Jülich-Aachen Research Alliance (JARA-HPC), RWTH Aachen University, Aachen 52056, Germany

Content

# Single-Crystal Diffraction on Bromomalonic Aldehyde

Table S1. Crystal data and structure refinement for bromomalonic aldehyde.

| Identification code | BMA (**B**romo**m**alonic **A**ldehyde) |
|---|---|
| Empirical formula | $C_3H_3BrO_2$ |
| Formula weight | 150.96 |
| Temperature | 100(2) K |
| Wavelength | 0.71073 Å |
| Crystal system | Orthorhombic |
| Space group | $Cmc2_1$ |
| $a$ (Å) | 6.31460(10) |
| $b$ (Å) | 10.7330(2) |
| $c$ (Å) | 6.38720(10) |
| $V$ (Å$^3$) | 432.890(13) |
| Z | 4 |
| Density (calculated, mg×mm$^{-3}$) | 2.316 |
| Absorption coefficient (mm$^{-1}$) | 9.334 |
| $F$(000) | 288 |
| Crystal size (mm) | 0.30×0.16×0.16 |
| Theta range for data collection (°) | 3.744 to 56.947 |
| Index ranges | $-14 \leq h \leq 13, -12 \leq k \leq 24, -14 \leq l \leq 14$ |
| Reflections collected | 17877 |
| Independent reflections | 3025 |
| $R_{int}$ | 0.0185 |
| Absorption correction | Semi-empirical from equivalents |
| Max. and min. transmission | 0.225 and 0.182 |
| Refinement method | Full-matrix least-squares on $F^2$ |
| Data / parameters | 3025 / 37 |
| Goodness-of-fit on $F^2$ | 1.065 |
| Final R indices [I>2σ(I)] | $R_1 = 0.0190$, $wR_2 = 0.0454$ |
| R indices (all data) | $R_1 = 0.0256$, $wR_2 = 0.0490$ |
| Absolute structure parameter | −0.001(8) |
| Largest diff. peak and hole (eÅ$^{-3}$) | 0.491 and −1.269 |

Table S2. Atomic coordinates for bromomalonic aldehyde

| | X | y | z |
|---|---|---|---|
| Br(1) | .5000 | .40465(2) | .54829(3) |
| O(1) | .5000 | .14233(12) | .77538(17) |
| O(2) | .5000 | .30282(11) | .08673(16) |
| C(1) | .5000 | .13756(13) | .58267(18) |
| C(2) | .5000 | .24171(11) | .44204(17) |
| C(3) | .5000 | .21702(12) | .23386(18) |
| H(2) | .5000 | .2678 | −.0309 |
| H(1) | .5000 | .0570 | .5212 |
| H(3) | .5000 | .1322 | .1914 |

Table S3. Bond lengths (in Å) and angles (°) for bromomalonic aldehyde

| | |
|---|---|
| Br(1)−C(2) | 1.8759(12) |
| O(1)−C(1) | 1.2319(16) |
| O(2)−C(3) | 1.3158(17) |
| O(2)−H(2) | 0.84 |
| C(1)−C(2) | 1.4340(17) |
| C(1)−H(1) | 0.95 |
| C(2)−C(3) | 1.3558(16) |
| C(3)−H(3) | 0.95 |
| C(3)−O(2)−H(2) | 109.0 |
| O(1)−C(1)−C(2) | 126.40(13) |
| O(1)−C(1)−H(1) | 116.8 |
| C(2)−C(1)−H(1) | 116.8 |
| C(3)−C(2)−C(1) | 117.51(11) |
| C(3)−C(2)−Br(1) | 122.48(9) |
| C(1)−C(2)−Br(1) | 120.00(9) |
| O(2)−C(3)−C(2) | 124.31(12) |
| O(2)−C(3)−H(3) | 117.8 |
| C(2)−C(3)−H(3) | 117.8 |

Table S4. Anisotropic displacement parameters for bromomalonic aldehyde. The anisotropic displacement factor exponent takes the form: $-2\pi^2 [ h^2 a^{*2} U_{11} +\ldots+ 2 h k a^* b^* U_{12} ]$.

| | $U_{11}$ | $U_{22}$ | $U_{33}$ | $U_{23}$ | $U_{13}$ | $U_{12}$ |
|---|---|---|---|---|---|---|
| Br(1) | .02125(5) | .01281(4) | .01279(4) | −.00187(5) | 0 | 0 |
| O(1) | .0264(5) | .0187(4) | .0108(3) | .0014(3) | 0 | 0 |
| O(2) | .0239(5) | .0182(3) | .0104(3) | .0013(2) | 0 | 0 |
| C(1) | .0226(5) | .0147(4) | .0113(4) | .0005(2) | 0 | 0 |
| C(2) | .0186(5) | .0134(3) | .0099(3) | .0001(2) | 0 | 0 |
| C(3) | .0191(5) | .0150(3) | .0101(3) | −.0007(3) | 0 | 0 |

# DFT calculations

*k*-point-convergence of total energies (ground state)

Table S5. *k*-point-convergence of total energies (ground state)

| Molecule | Level of Theory | Structure | *k*-point mesh | Energy (eV) |
|---|---|---|---|---|
| Urea | PBE+D3(BJ) | Fully optimized | 10×10×11 | −98.27743389 |
| | | 10×10×11-structure | 11×11×12 | −98.27747547 |
| Urea | PBE+D3M(BJ) | Fully optimized | 10×10×12 | −98.29783825 |
| | | 10×10×12-structure | 11×11×13 | −98.29779450 |
| Naphthalene | PBE+D3(BJ) | Fully optimized | 8×9×8 | −241.93176170 |
| | | 8×9×8-structure | 9×10×9 | −241.93176994 |
| Naphthalene | PBE+D3M(BJ) | Fully optimized | 8×10×8 | −242.30511059 |
| | | 8×10×8-structure | 9×11×9 | −242.30511732 |
| Pentachloropyridine | PBE+D3(BJ) | Fully optimized | 10×10×4 | −130.47695112 |
| | | 10×10×4-structure | 11×11×5 | −130.47693657 |
| Pentachloropyridine | PBE+D3M(BJ) | Fully optimized | 10×10×4 | −131.31733222 |
| | | 10×10×4-structure | 11×11×5 | −131.31733075 |
| Bromomalonic Aldehyde | PBE+D3(BJ) | Fully optimized | 8×5×8 | −214.07782876 |
| | | 8×5×8-structure | 9×6×9 | −214.07778797 |
| Bromomalonic Aldehyde | PBE+D3M(BJ) | Fully optimized | 9×6×8 | −214.67584375 |
| | | 9×6×8-structure | 10×7×9 | −214.67587341 |

*q*-point convergence of ADPs (PBE+D3M(BJ) level)

The following q-point meshes were at least used to calculate the ADPs:

Urea: 60×60×66

Naphthalene: 32×36×28

Pentachloropyridine: 70×70×28

Bromomalonic aldehyde: 42×42×42

Table S6. *q*-point convergence of ADPs (PBE+D3M(BJ) level)

| **Molecule** | Structure | *q*-point mesh | Frequency cutoff | Temperature (K) | Atom Type | $U_{eq}$ (Å$^2$) |
|---|---|---|---|---|---|---|
| Urea | 123K-Volume | 40×40×44 | 0.1 THz | 123 K | H1 | 3.774e−02 |
|  | 123K-Volume | 50×50×55 | 0.1 THz | 123 K | H1 | 3.805e−02 |
|  | 123K-Volume | 60×60×66 | 0.1 THz | 123 K | H1 | 3.771e−02 |
|  | 123K-Volume | 70×70×77 | 0.1 THz | 123 K | H1 | 3.779e−02 |
| Naphthalene | 295K-Volume | 30×34×26 | 0.1 THz | 295 K | H1 | 1.464e−01 |
|  | 295K-Volume | 32×36×28 | 0.1 THz | 295 K | H1 | 1.464e−01 |
|  | 295K-Volume | 35×39×31 | 0.1 THz | 295K | H1 | 1.463e−01 |
| Pentachloropyridine | 300 K-volume | 68×68×26 | 0.13 THz | 300 K | Cl5 | 6.653e−02 |
|  | 300 K-volume | 70×70×28 | 0.13 THz | 300 K | Cl5 | 6.657e−02 |
| Bromomalonic aldehyde | 100 K-volume | 40×40×40 | 0.1 THz | 100 K | H1 | 3.720e−02 |
|  | 100 K-volume | 42×42×42 | 0.1 THz | 100 K | H1 | 3.720e−02 |

Influence of the supercell size on the ADPs

We checked the influence of the supercell size on the ADPs calculated in the harmonic approximation at the PBE+D3M(BJ) level of theory. The $q$-point meshes were at least as large as mentioned in the last section.

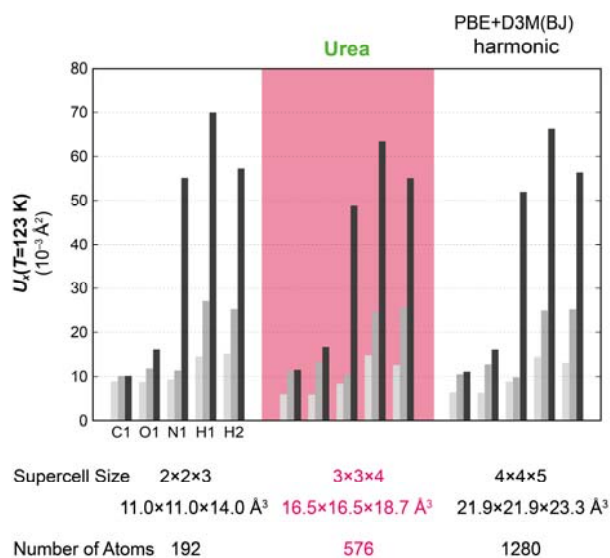

Figure S1. Influence of the supercell size on the calculated ADPs for urea in the harmonic approximation and at the PBE+D3M(BJ) level of theory. All calculations shown in the main text were performed with the 3×3×4 supercell of the conventional cell. Performing all our calculations with the 4×4×5 supercell is computationally not feasible.

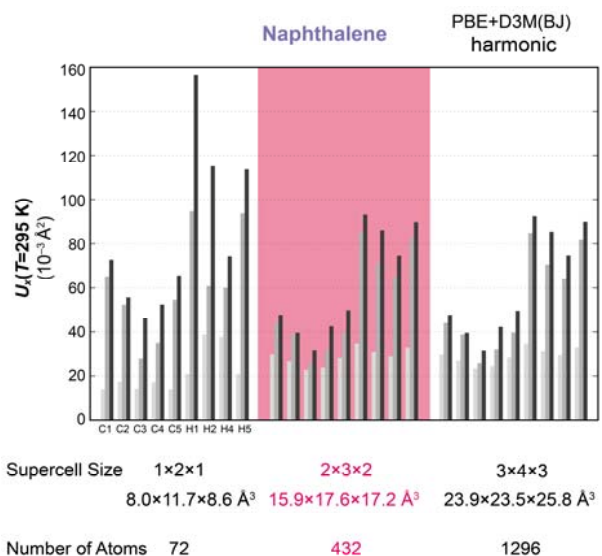

Figure S2. Influence of the supercell size on the calculated ADPs for naphthalene in the harmonic approximation and at the PBE+D3M(BJ) level of theory. All calculations shown in the main text were performed with the 2×3×2 supercell of the conventional cell. Performing all our calculations with the 3×4×3 supercell is computationally not feasible.

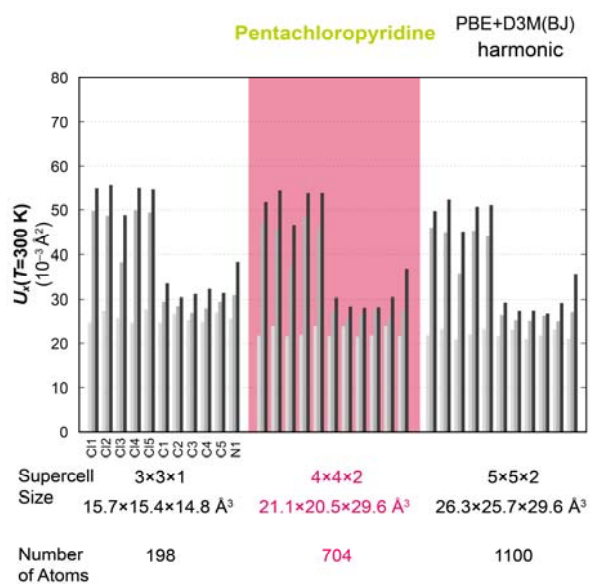

Figure S3. Influence of the supercell size on the calculated ADPs for pentachloropyridine in the harmonic approximation and at the PBE+D3M(BJ) level of theory. All calculations shown in the main text were performed with the 4×4×2 supercell of the conventional cell. Performing all our calculations with the 5×5×2 supercell is computationally not feasible.

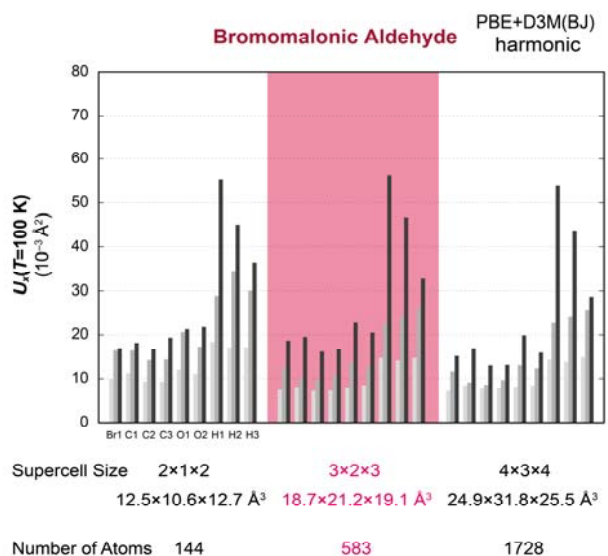

Figure S4. Influence of the supercell size on the calculated ADPs for bromomalonic aldehyde in the harmonic approximation and at the PBE+D3M(BJ) level of theory. All calculations shown in the main text were performed with the 3×2×3 supercell of the conventional cell. Performing all our calculations with the 4×3×4 supercell is computationally not feasible.

# Thermal Expansion
Urea

Table S7. Temperature-dependent volumes per Formula unit *V*/*Z* for urea from the quasi-harmonic approximation at two levels of theory. The temperatures at which we calculated ADPs are marked by a bold font.

| T(K) | V/Z (Å$^3$) | |
| --- | --- | --- |
|  | PBE+D3(BJ) | PBE+D3M(BJ) |
| 0 | 73.38 | 73.33 |
| **12** |  | **73.33** |
| 25 | 73.39 | 73.34 |
| 50 | 73.46 | 73.41 |
| **60** |  | **73.45** |
| 75 | 73.60 | 73.54 |
| 100 | 73.79 | 73.73 |
| **123** |  | **73.93** |
| 125 | 74.01 | 73.95 |
| 150 | 74.27 | 74.20 |
| 175 | 74.55 | 74.48 |
| 200 | 74.86 | 74.79 |
| 225 | 75.19 | 75.12 |
| 250 | 75.55 | 75.48 |
| 275 | 75.94 | 75.87 |
| 300 | 76.35 | 76.29 |

Table S8. Temperature-dependent lattice parameters for urea from the quasi-harmonic approximation at two levels of theory. The temperatures at which we calculated ADPs are marked by a bold font.

| *T* (K) | PBE+D3(BJ) | | PBE+D3M(BJ) | |
| --- | --- | --- | --- | --- |
|  | *a* (Å) | *c* (Å) | *a* (Å) | *c* (Å) |
| 0 | 5.598 | 4.683 | 5.599 | 4.679 |
| **12** |  |  | **5.599** | **4.679** |
| 25 | 5.599 | 4.683 | 5.599 | 4.679 |
| 50 | 5.601 | 4.683 | 5.601 | 4.680 |
| **60** |  |  | **5.603** | **4.680** |
| 75 | 5.606 | 4.684 | 5.606 | 4.680 |
| 100 | 5.613 | 4.684 | 5.613 | 4.681 |
| **123** |  |  | **5.620** | **4.682** |
| 125 | 5.621 | 4.685 | 5.621 | 4.682 |
| 150 | 5.630 | 4.687 | 5.630 | 4.683 |
| 175 | 5.640 | 4.688 | 5.639 | 4.684 |
| 200 | 5.650 | 4.690 | 5.650 | 4.686 |
| 225 | 5.662 | 4.691 | 5.661 | 4.687 |
| 250 | 5.674 | 4.693 | 5.675 | 4.688 |
| 275 | 5.688 | 4.694 | 5.688 | 4.690 |
| 300 | 5.703 | 4.694 | 5.703 | 4.692 |

Bromomalonic aldehyde

Table S9. Temperature-dependent volumes per formula unit *V*/*Z* for bromomalonic aldehyde from the quasi-harmonic approximation at two levels of theory. The temperatures at which we calculated ADPs are marked by a bold font.

| *T* (K) | *V*/*Z* ($Å^3$) | |
| --- | --- | --- |
| | PBE+D3(BJ) | PBE+D3M(BJ) |
| 0 | 112.52 | 109.01 |
| 25 | 112.60 | 109.09 |
| 50 | 112.93 | 109.38 |
| 75 | 113.38 | 109.81 |
| **100** | 113.90 | **110.30** |
| 125 | 114.46 | 110.84 |
| 150 | 115.06 | 111.42 |
| 175 | 115.69 | 112.02 |
| 200 | 116.34 | 112.66 |
| 225 | 117.02 | 113.32 |
| 250 | 117.72 | 114.02 |
| 275 | 118.44 | 114.74 |
| 300 | 119.19 | 115.48 |

Table S10. Temperature-dependent lattice parameters for bromomalonic aldehyde from the quasi-harmonic approximation at two levels of theory. The temperatures at which we calculated ADPs are marked by a bold font.

| *T* (K) | PBE+D3(BJ) | | | PBE+D3M(BJ) | | |
| --- | --- | --- | --- | --- | --- | --- |
| | *a* (Å) | *b* (Å) | *c* (Å) | *a* (Å) | *b* (Å) | *c* (Å) |
| 0 | 6.501 | 10.827 | 6.394 | 6.372 | 10.724 | 6.381 |
| 25 | 6.503 | 10.831 | 6.394 | 6.376 | 10.725 | 6.381 |
| 50 | 6.515 | 10.842 | 6.394 | 6.387 | 10.733 | 6.383 |
| 75 | 6.532 | 10.856 | 6.396 | 6.400 | 10.750 | 6.384 |
| 100 | 6.551 | 10.871 | 6.397 | 6.419 | 10.766 | 6.384 |
| 125 | 6.570 | 10.890 | 6.399 | 6.440 | 10.781 | 6.386 |
| 150 | 6.593 | 10.909 | 6.399 | 6.459 | 10.802 | 6.388 |
| 175 | 6.615 | 10.928 | 6.402 | 6.481 | 10.820 | 6.390 |
| 200 | 6.637 | 10.950 | 6.403 | 6.504 | 10.842 | 6.391 |
| 225 | 6.661 | 10.975 | 6.403 | 6.528 | 10.864 | 6.391 |
| 250 | 6.687 | 10.994 | 6.405 | 6.554 | 10.885 | 6.393 |
| 275 | 6.712 | 11.018 | 6.406 | 6.582 | 10.904 | 6.395 |
| 300 | 6.736 | 11.042 | 6.409 | 6.610 | 10.925 | 6.397 |

Pentachloropyridine

Table S11. Temperature-dependent volumes per formula unit *V*/*Z* for pentachloropyridine from the quasi-harmonic approximation at two levels of theory. The temperatures at which we calculated ADPs are marked by a bold font.

| T(K) | *V*/*Z* (Å$^3$) | |
|---|---|---|
| | PBE+D3(BJ) | PBE+D3M(BJ) |
| 0 | 210.95 | 202.04 |
| 25 | 211.16 | 202.20 |
| 50 | 211.83 | 202.78 |
| 75 | 212.71 | 203.56 |
| **100** | 213.69 | **204.44** |
| 125 | 214.76 | 205.41 |
| **150** | 215.91 | **206.44** |
| 175 | 217.13 | 207.54 |
| **200** | 218.43 | **208.70** |
| 225 | 219.81 | 209.94 |
| **250** | 221.28 | **211.25** |
| 275 | 222.85 | 212.63 |
| **300** | 224.53 | **214.11** |

Table S12. Temperature-dependent lattice parameters for pentachloropyridine from the quasi-harmonic approximation at two levels of theory. The temperatures at which we calculated ADPs are marked by a bold font.

| T(K) | PBE+D3(BJ) | | | PBE+D3M(BJ) | | |
|---|---|---|---|---|---|---|
| | *a* (Å) | *b* (Å) | *c* (Å) | *a* (Å) | *b* (Å) | *c* (Å) |
| 0 | 5.376 | 5.216 | 15.301 | 5.307 | 5.162 | 14.972 |
| 25 | 5.378 | 5.218 | 15.305 | 5.309 | 5.162 | 14.978 |
| 50 | 5.383 | 5.220 | 15.336 | 5.316 | 5.164 | 15.001 |
| 75 | 5.386 | 5.222 | 15.388 | 5.321 | 5.170 | 15.026 |
| 100 | 5.400 | 5.235 | 15.382 | 5.327 | 5.176 | 15.059 |
| 125 | 5.403 | 5.235 | 15.450 | 5.336 | 5.182 | 15.090 |
| 150 | 5.417 | 5.247 | 15.459 | 5.345 | 5.184 | 15.139 |
| 175 | 5.430 | 5.258 | 15.483 | 5.352 | 5.194 | 15.172 |
| 200 | 5.428 | 5.254 | 15.591 | 5.364 | 5.195 | 15.230 |
| 225 | 5.441 | 5.265 | 15.624 | 5.371 | 5.210 | 15.252 |
| 250 | 5.454 | 5.278 | 15.658 | 5.380 | 5.213 | 15.322 |
| 275 | 5.467 | 5.288 | 15.702 | 5.388 | 5.229 | 15.353 |
| 300 | 5.473 | 5.298 | 15.775 | 5.403 | 5.230 | 15.420 |

Naphthalene

Table S13. Temperature-dependent volumes per formula unit *V*/Z for naphthalene from the quasi-harmonic approximation at two levels of theory. The temperatures at which we calculated ADPs are marked by a bold font.

| T(K) | *V*/Z (Å$^3$) | |
|---|---|---|
| | PBE+D3(BJ) | PBE+D3M(BJ) |
| **0** | 174.54 | 173.06 |
| 5 | 174.54 | 173.06 |
| 25 | 174.68 | 173.2 |
| 50 | 175.24 | 173.77 |
| 75 | 176.04 | 174.58 |
| **80** | 176.22 | 174.76 |
| 100 | 176.99 | 175.55 |
| 125 | 178.07 | 176.63 |
| **150** | 179.26 | 177.83 |
| 175 | 180.57 | 179.16 |
| 200 | 182.02 | 180.62 |
| **220** | 183.30 | 181.89 |
| 225 | 183.63 | 182.23 |
| 250 | 185.43 | 184.02 |
| 275 | 187.46 | 186.03 |
| **295** | 189.29 | 187.83 |
| 300 | 189.78 | 188.31 |

Table S14. Temperature-dependent lattice parameters for naphthalene from the quasi-harmonic approximation at two levels of theory. The temperatures at which we calculated ADPs are marked by a bold font.

| T(K) | PBE+D3(BJ) | | | PBE+D3M(BJ) | | |
|---|---|---|---|---|---|---|
| | *a* (Å) | *b* (Å) | c (Å) | *a* (Å) | *b* (Å) | *c* (Å) |
| 0 | 8.155 | 5.960 | 8.678 | 8.117 | 5.949 | 8.687 |
| 25 | 8.159 | 5.961 | 8.679 | 8.115 | 5.953 | 8.688 |
| 50 | 8.173 | 5.964 | 8.684 | 8.134 | 5.955 | 8.703 |
| 75 | 8.182 | 5.976 | 8.696 | 8.148 | 5.963 | 8.705 |
| 100 | 8.203 | 5.984 | 8.720 | 8.168 | 5.972 | 8.720 |
| 125 | 8.226 | 5.992 | 8.728 | 8.186 | 5.984 | 8.736 |
| 150 | 8.246 | 6.006 | 8.749 | 8.212 | 5.995 | 8.770 |
| 175 | 8.272 | 6.016 | 8.773 | 8.234 | 6.007 | 8.782 |
| 200 | 8.299 | 6.031 | 8.790 | 8.264 | 6.020 | 8.809 |
| 225 | 8.326 | 6.047 | 8.817 | 8.292 | 6.033 | 8.838 |
| 250 | 8.380 | 6.049 | 8.850 | 8.328 | 6.050 | 8.863 |
| 275 | 8.415 | 6.069 | 8.875 | 8.361 | 6.070 | 8.888 |
| 300 | 8.447 | 6.096 | 8.927 | 8.408 | 6.083 | 8.930 |

# Irreducible representations of IR- and Raman-active modes

The following irreducible representations have been extracted by the program SAM on the Bilbao crystallographic server. [1]

Urea
$M = 8A_1 + 3A_2 + 3B_1 + 8B_2 + 13E$

IR active: $7B_2 + 12E$

Raman active: $8A_1 + 3B_1 + 7B_2 + 12E$

Bromomalonic aldehyde
$M = 18A_1 + 9A_2 + 9B_1 + 18B_2$

IR active: $17A_1 + 8B_1 + 17B_2$

Raman active: $17A_1 + 9A_2 + 8B_1 + 17B_2$

Pentachloropyridine
$M = 33A' + 33A''$

IR active: $31A' + 32A''$

Raman active: $31A' + 32A''$

Naphthalene
$M = 27A_g + 27A_u + 27B_g + 27B_u$

IR active: $26A_u + 25B_u$

Raman active: $27A_g + 27B_g$

# Mulliken Symbols of irreducible representations for frequencies at Γ at PBE+D3M(BJ) level of theory (ground state)

Urea

Table S15. Irreducible representations of urea's frequencies at Γ at PBE+D3M(BJ) level of theory (ground state). All frequencies are given in cm$^{-1}$.

| | |
|---|---|
| E | -0.37292424 |
| B$_2$ | 0.27895337 |
| B$_1$ | 54.1661356 |
| A$_2$ | 83.5426578 |
| E | 113.165532 |
| A$_1$ | 134.543507 |
| E | 170.881506 |
| E | 218.336562 |
| B$_1$ | 395.851847 |
| E | 529.635524 |
| A$_1$ | 551.692637 |
| E | 556.798059 |
| B$_2$ | 575.415866 |
| A$_2$ | 603.071249 |
| B$_1$ | 652.126194 |
| A$_2$ | 690.598856 |
| E | 753.788214 |
| E | 788.733864 |
| B$_2$ | 1013.85019 |
| A$_1$ | 1017.71614 |
| E | 1073.04735 |
| B$_2$ | 1128.03552 |
| A$_1$ | 1168.91444 |
| E | 1478.92945 |
| A$_1$ | 1493.85572 |
| B$_2$ | 1560.039 |
| E | 1609.98578 |
| A$_1$ | 1622.35615 |
| B$_2$ | 1658.40174 |
| E | 3308.36275 |
| A$_1$ | 3339.56556 |
| B$_2$ | 3348.6812 |
| E | 3457.13177 |
| A$_1$ | 3458.86635 |
| B$_2$ | 3490.64349 |

Bromomalonic aldehyde

Table S16. Irreducible representations of bromomalonic aldehyde's frequencies at Γ at PBE+D3M(BJ) level of theory (ground state). All frequencies are given in cm$^{-1}$.

| Irrep | Frequency | Irrep | Frequency |
|---|---|---|---|
| $B_1$ | -1.4818054 | $B_2$ | 1461.03769 |
| $A_1$ | -0.74507028 | $A_1$ | 1525.20786 |
| $B_2$ | -0.40314289 | $B_2$ | 1544.51339 |
| $A_2$ | 48.4548994 | $A_1$ | 1651.69501 |
| $B_2$ | 61.8791619 | $B_2$ | 1664.90887 |
| $A_1$ | 78.9539917 | $A_1$ | 2498.36596 |
| $B_1$ | 90.1181099 | $B_2$ | 2504.12091 |
| $A_2$ | 98.1942207 | $B_2$ | 2914.08746 |
| $B_2$ | 116.257246 | $A_1$ | 2920.00683 |
| $A_1$ | 123.288618 | $A_1$ | 3081.4765 |
| $B_1$ | 128.20948 | $B_2$ | 3082.8149 |
| $A_2$ | 134.54124 | | |
| $A_2$ | 171.004243 | | |
| $B_1$ | 189.715883 | | |
| $A_1$ | 194.80027 | | |
| $B_2$ | 201.836073 | | |
| $B_2$ | 290.418741 | | |
| $A_1$ | 294.5026 | | |
| $B_1$ | 389.569411 | | |
| $A_2$ | 391.135512 | | |
| $B_1$ | 430.829885 | | |
| $A_2$ | 432.43437 | | |
| $B_2$ | 515.147053 | | |
| $A_1$ | 516.847998 | | |
| $B_2$ | 543.478819 | | |
| $A_1$ | 548.653273 | | |
| $A_1$ | 698.380343 | | |
| $B_2$ | 716.309581 | | |
| $B_1$ | 874.915849 | | |
| $A_2$ | 879.569462 | | |
| $A_2$ | 952.361353 | | |
| $B_1$ | 954.475224 | | |
| $A_2$ | 1055.26072 | | |
| $B_1$ | 1057.01631 | | |
| $A_1$ | 1200.41357 | | |
| $B_2$ | 1203.58626 | | |
| $A_1$ | 1251.00375 | | |
| $B_2$ | 1261.98688 | | |
| $A_1$ | 1299.39978 | | |
| $B_2$ | 1301.21432 | | |
| $A_1$ | 1395.1634 | | |
| $B_2$ | 1396.55894 | | |
| $A_1$ | 1454.58284 | | |

Pentachloropyridine

Table S17. Irreducible representations of bromomalonic aldehyde's frequencies at Γ at PBE+D3M(BJ) level of theory (ground state). All frequencies are given in cm$^{-1}$.

| | | | | |
|---|---|---|---|---|
| A" | -0.32453367 | | A' | 572.710018 |
| A' | -0.30184434 | | A" | 572.925134 |
| A' | -0.10032757 | | A" | 609.276896 |
| A' | 35.257555 | | A' | 610.348651 |
| A" | 35.9599542 | | A' | 676.010461 |
| A" | 39.7828209 | | A" | 683.552378 |
| A' | 46.0900971 | | A' | 715.781734 |
| A" | 52.3307848 | | A" | 716.151048 |
| A" | 58.599512 | | A' | 811.691709 |
| A' | 60.7973692 | | A" | 811.710486 |
| A' | 88.5197351 | | A" | 869.060546 |
| A" | 92.9002886 | | A' | 869.579931 |
| A" | 95.2507256 | | A' | 1066.96748 |
| A' | 100.486071 | | A" | 1067.30117 |
| A' | 107.827485 | | A" | 1204.50783 |
| A" | 110.58474 | | A' | 1204.60609 |
| A" | 175.071213 | | A' | 1287.38504 |
| A' | 176.208028 | | A" | 1292.41454 |
| A' | 197.958464 | | A' | 1308.4252 |
| A" | 201.64394 | | A" | 1308.73874 |
| A" | 210.120831 | | A' | 1310.08749 |
| A' | 215.495408 | | A" | 1323.76345 |
| A' | 219.552689 | | A' | 1481.62733 |
| A" | 219.842196 | | A" | 1482.463 |
| A' | 222.202438 | | A' | 1503.46009 |
| A" | 223.497893 | | A" | 1504.42511 |
| A' | 280.877639 | | | |
| A" | 281.078997 | | | |
| A" | 326.883939 | | | |
| A' | 327.899097 | | | |
| A" | 346.034226 | | | |
| A' | 349.813508 | | | |
| A' | 363.974338 | | | |
| A" | 365.819776 | | | |
| A" | 382.995718 | | | |
| A' | 383.254443 | | | |
| A' | 543.795632 | | | |
| A" | 544.300904 | | | |
| A" | 563.268436 | | | |
| A' | 563.482602 | | | |

Naphthalene

Table S18. Irreducible representations of naphthalene's frequencies at Γ at PBE+D3M(BJ) level of theory (ground state). All frequencies are given in cm$^{-1}$.

| | | | | | |
|---|---|---|---|---|---|
| $A_u$ | -0.27891929 | $A_g$ | 770.954964 | $A_u$ | 1257.13962 |
| $B_u$ | -0.23506 | $B_u$ | 772.212421 | $B_u$ | 1261.76497 |
| $B_u$ | -0.05152966 | $B_g$ | 775.164311 | $A_u$ | 1374.4355 |
| $B_g$ | 58.183236 | $A_u$ | 786.504568 | $B_u$ | 1374.52876 |
| $A_u$ | 59.67305 | $A_u$ | 792.080476 | $A_u$ | 1380.91251 |
| $A_g$ | 67.0590033 | $B_u$ | 794.314973 | $B_u$ | 1382.41229 |
| $B_u$ | 81.2363965 | $A_u$ | 839.85382 | $B_g$ | 1392.6672 |
| $B_g$ | 86.8330028 | $B_u$ | 844.386217 | $A_g$ | 1393.72926 |
| $A_g$ | 93.613146 | $A_g$ | 877.688757 | $B_g$ | 1442.69966 |
| $A_u$ | 115.208561 | $B_g$ | 892.305877 | $B_g$ | 1447.60013 |
| $A_g$ | 128.396359 | $A_g$ | 924.994927 | $A_g$ | 1449.40285 |
| $B_g$ | 148.562305 | $B_g$ | 929.129717 | $A_g$ | 1451.24522 |
| $B_u$ | 177.691515 | $B_g$ | 941.791837 | $B_u$ | 1506.51433 |
| $A_u$ | 197.084278 | $A_g$ | 946.066643 | $A_u$ | 1509.0264 |
| $A_u$ | 217.76328 | $B_u$ | 955.043174 | $A_g$ | 1566.55145 |
| $B_u$ | 219.070694 | $A_u$ | 955.362757 | $B_g$ | 1569.23792 |
| $B_u$ | 357.031864 | $A_u$ | 973.389811 | $A_u$ | 1590.0595 |
| $A_u$ | 360.367189 | $A_g$ | 978.67751 | $B_u$ | 1590.51542 |
| $B_g$ | 386.642707 | $B_u$ | 979.834078 | $B_g$ | 1622.30937 |
| $A_g$ | 392.289937 | $B_g$ | 982.65101 | $A_g$ | 1622.66316 |
| $A_g$ | 463.673418 | $B_u$ | 1016.8512 | $A_u$ | 3093.43738 |
| $A_u$ | 468.519361 | $A_u$ | 1017.55789 | $A_g$ | 3094.44146 |
| $B_g$ | 469.49307 | $A_g$ | 1022.33473 | $B_u$ | 3095.29527 |
| $B_u$ | 480.86276 | $B_g$ | 1026.074 | $B_g$ | 3096.37515 |
| $B_g$ | 505.642228 | $A_u$ | 1115.32402 | $B_u$ | 3106.90647 |
| $A_g$ | 507.149349 | $B_u$ | 1115.46302 | $A_u$ | 3107.03317 |
| $B_g$ | 511.361825 | $A_g$ | 1136.15214 | $A_g$ | 3107.39286 |
| $A_g$ | 512.506504 | $B_g$ | 1139.62685 | $B_g$ | 3107.42919 |
| $B_u$ | 614.729282 | $A_u$ | 1140.0611 | $B_g$ | 3117.18076 |
| $A_u$ | 616.733947 | $A_g$ | 1142.25252 | $B_u$ | 3117.84807 |
| $A_u$ | 622.738166 | $B_u$ | 1148.33514 | $A_g$ | 3118.22445 |
| $B_u$ | 627.781662 | $B_g$ | 1150.91985 | $A_u$ | 3118.28007 |
| $A_g$ | 715.246706 | $A_u$ | 1218.93194 | $B_u$ | 3120.79645 |
| $B_g$ | 717.638513 | $B_u$ | 1219.28385 | $A_u$ | 3122.32394 |
| $B_g$ | 766.158514 | $B_g$ | 1235.88958 | $B_g$ | 3123.06202 |
| $A_g$ | 769.018849 | $A_g$ | 1239.042 | $A_g$ | 3124.46413 |

Contribution of each atom to the vibrational states at the Γ point

This is done similar to references 2 and 3.

Contribution of atom *i* to vibrational state of mode $m = \frac{\sum_\alpha [e_m(i\alpha)]^* e_m(i\alpha)}{\sum_{k,\alpha}[e_m(k\alpha)]^* e_m(k\alpha)} = \frac{\sum_\alpha [e_m(i\alpha)]^* e_m(i\alpha)}{1}$

$e_m(i\alpha)$ is the eigenvector of atom *i* in mode *m* along the Cartesian direction α

We show the sums over all crystallographic equivalent atoms.

Urea

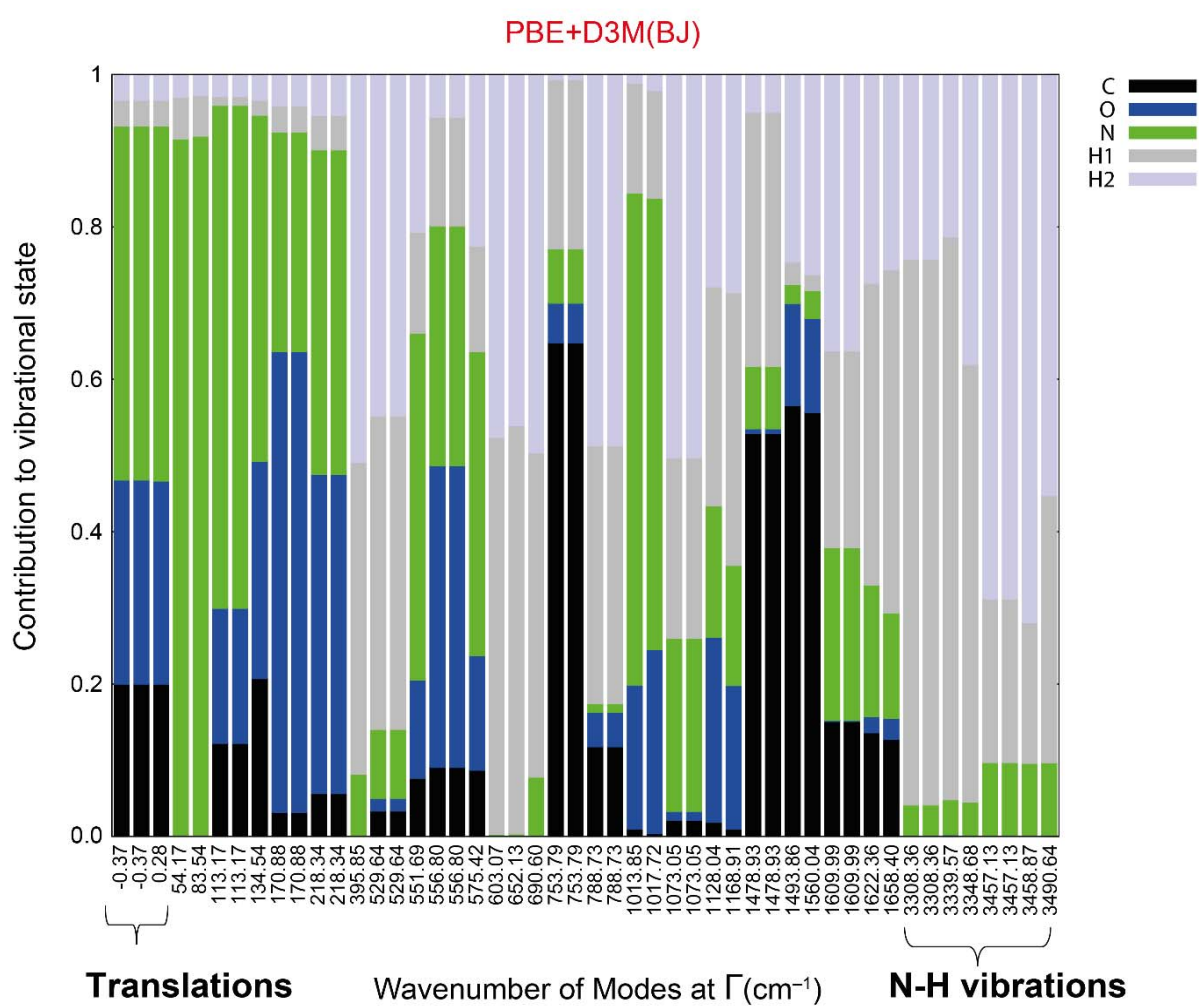

Fig S5. Contribution of each atom to the vibrational states at the Γ point for urea. The atoms are numbered as in ref 4.

Pentachloropyridine

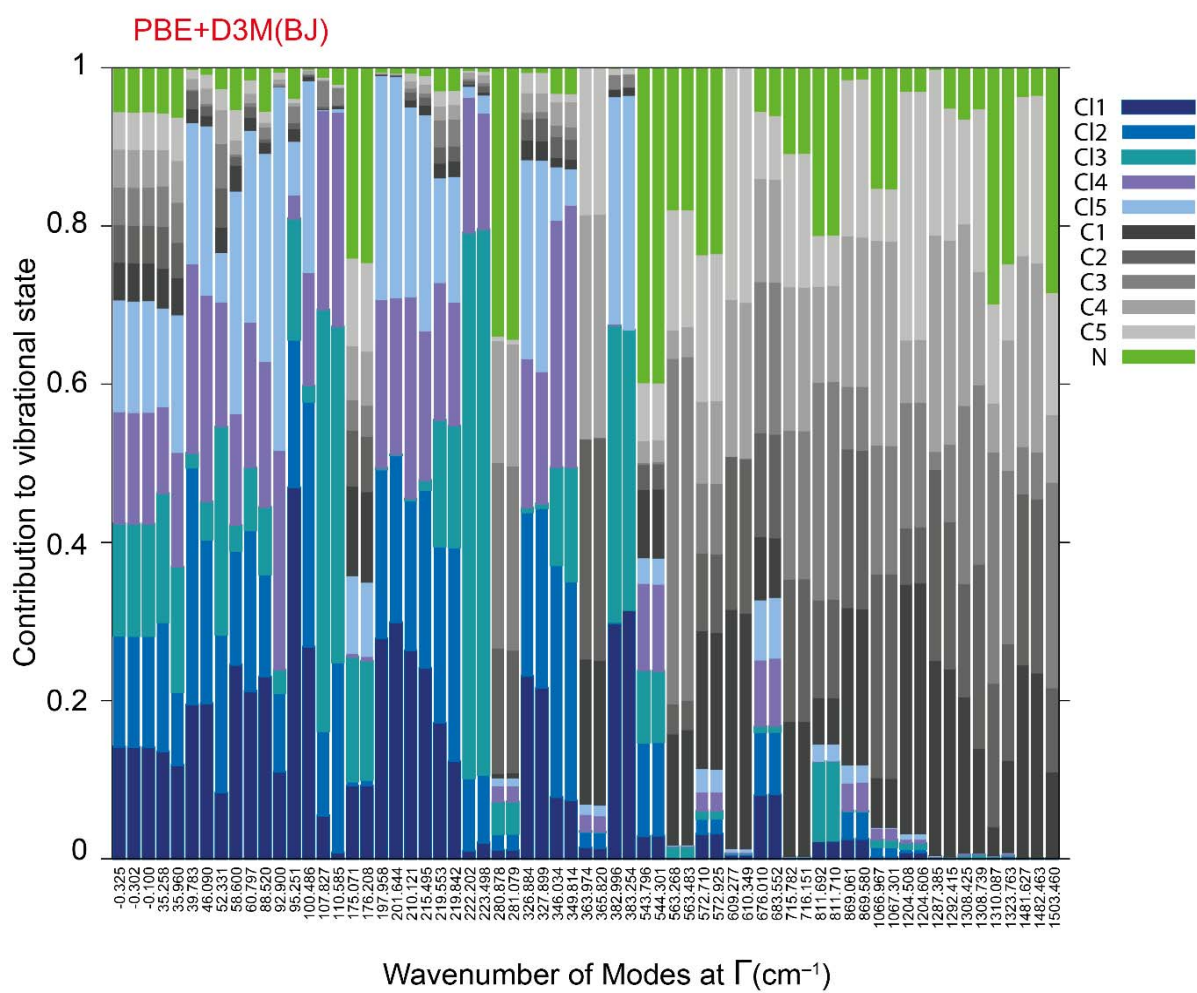

Fig S6. Contribution of each atom to the vibrational states at the Γ point for pentachloropyridine. The atoms are numbered as in reference 5 .

Bromomalonic aldehyde

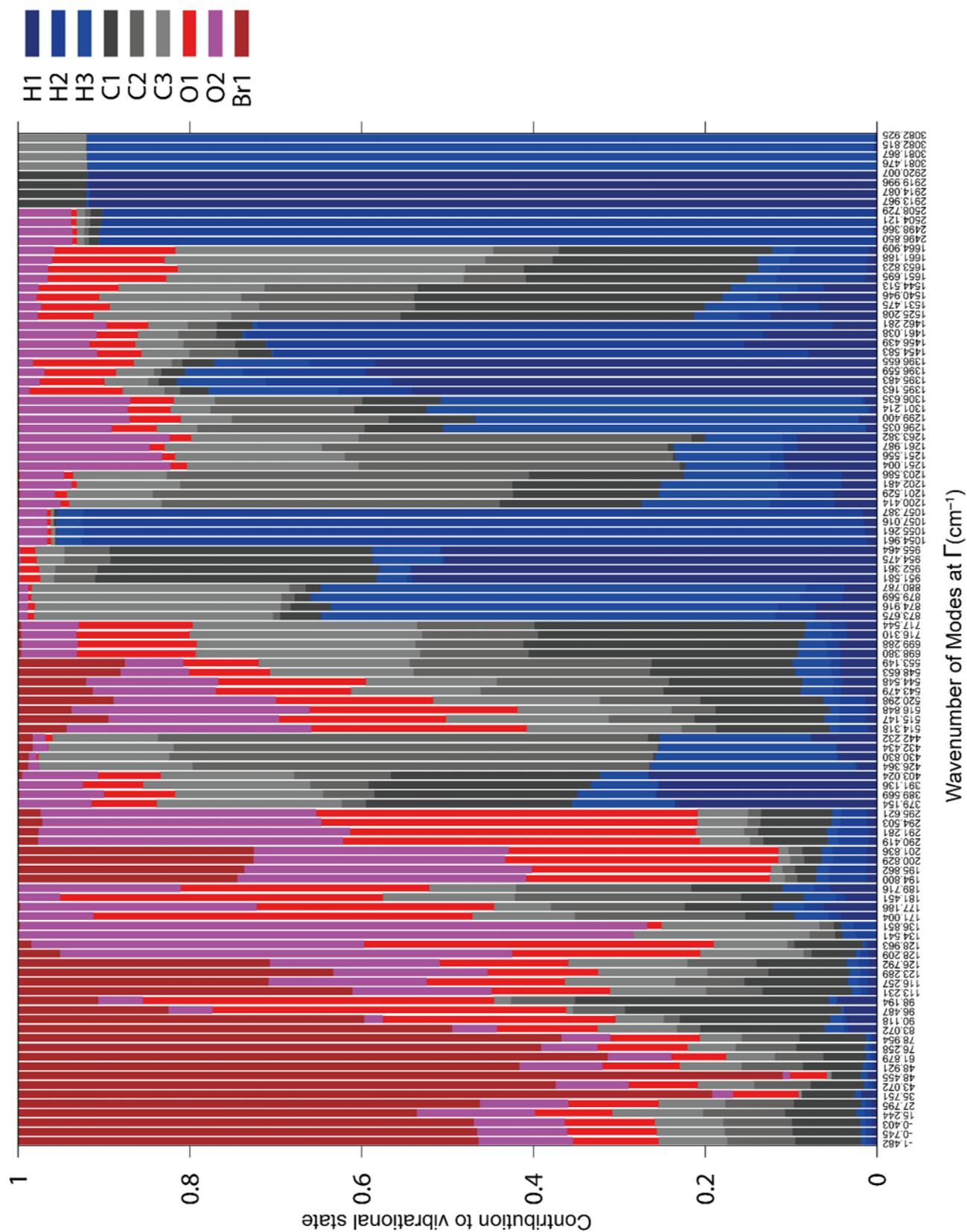

Fig S7. Contribution of each atom to the vibrational states at the Γ point for bromomalonic aldehyde. The atoms are numbered as in ref 6.

Napthalene

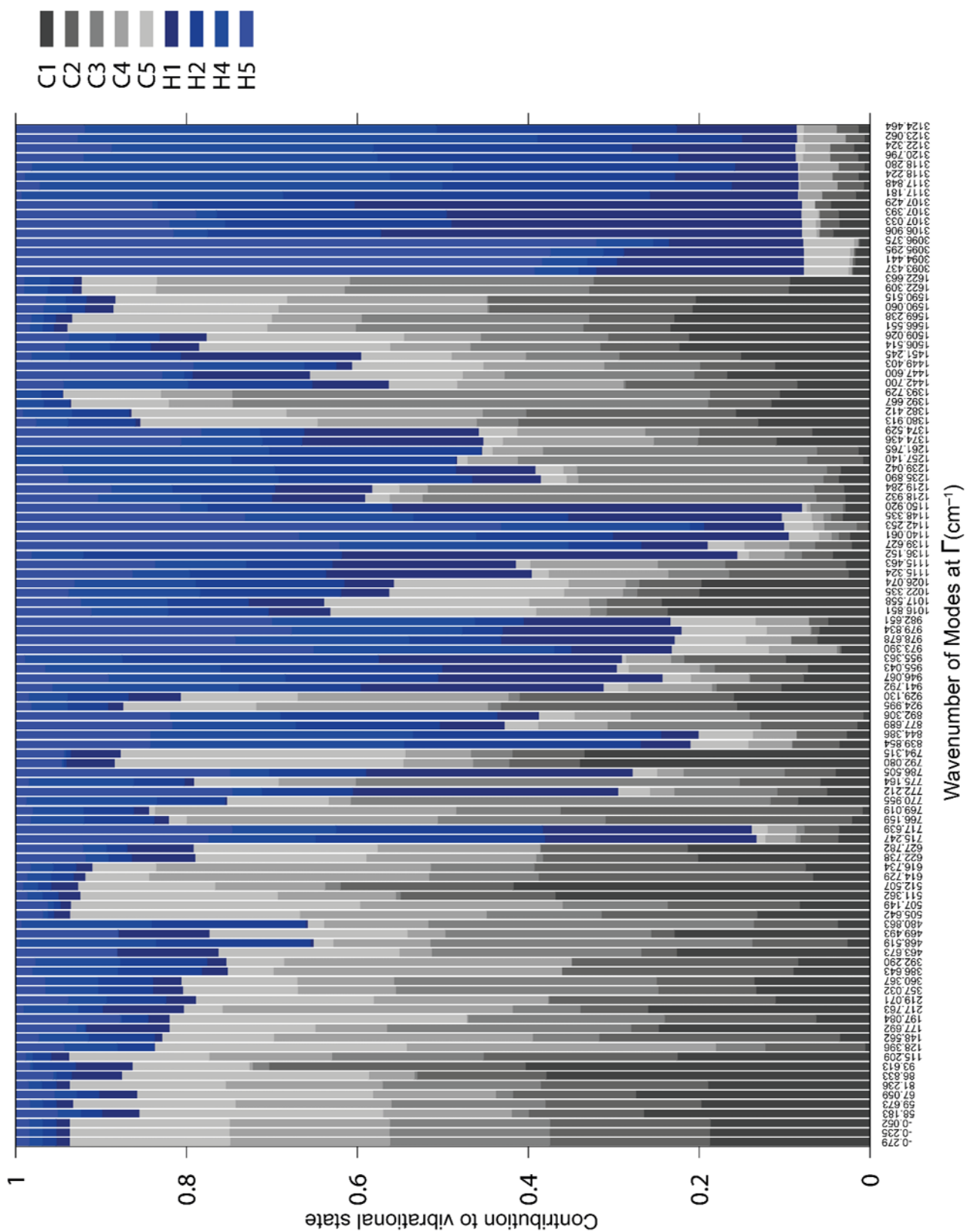

Fig S8. Contribution of each atom to the vibrational states at the Γ point napthalene. The atoms are numbered as in ref 7.

# Normal-mode anharmonicity

Urea

Table S19. Urea's frequencies from Phonopy (harmonic frequencies) and from the Frozen-Phonon Calculations (harmonic and anharmonic Frequencies) at the Γ-point at PBE+D3M(BJ) level of theory (123 K-volume). All frequencies are given in cm$^{-1}$.

| Mode | Mulliken Symbol | Harmonic Frequency (Phonopy) | Harmonic Frequency (Frozen Phonon) | First Anharmonic Frequency (Frozen Phonon) |
|---|---|---|---|---|
| 4  | B$_1$ | 54.6  | 57.3  | 76.6  |
| 5  | A$_2$ | 67.7  | 69.8  | 86.3  |
| 6  | E    | 102.1 | 103.3 | 106.5 |
| 7  | E    | 102.1 | 103.3 | 106.5 |
| 8  | A$_1$ | 119.2 | 119.6 | 119.2 |
| 9  | E    | 157.4 | 158.4 | 161.6 |
| 10 | E    | 157.4 | 158.4 | 161.6 |
| 11 | E    | 190.6 | 191.1 | 193.8 |
| 12 | E    | 190.6 | 191.1 | 193.8 |
| 13 | B$_1$ | 386.8 | 384.8 | 489.0 |
| 14 | E    | 515.8 | 517.4 | 617.8 |
| 15 | E    | 515.8 | 517.4 | 617.8 |
| 16 | A$_1$ | 545.6 | 546.1 | 548.5 |
| 17 | E    | 552.7 | 552.8 | 555.3 |
| 18 | E    | 552.7 | 552.8 | 555.3 |
| 19 | B$_2$ | 563.9 | 564.6 | 568.1 |
| 20 | A$_2$ | 593.7 | 594.4 | 639.9 |
| 21 | B$_1$ | 633.0 | 633.8 | 672.7 |
| 22 | A$_2$ | 665.9 | 666.8 | 711.5 |
| 23 | E    | 749.8 | 750.1 | 755.6 |
| 24 | E    | 749.8 | 750.1 | 755.6 |
| 25 | E    | 782.8 | 783.5 | 809.9 |
| 26 | E    | 782.8 | 783.5 | 809.9 |
| 27 | B$_2$ | 1005.0 | 1005.1 | 1006.7 |
| 28 | A$_1$ | 1009.0 | 1009.0 | 1008.1 |
| 29 | E    | 1063.5 | 1064.2 | 1076.2 |
| 30 | E    | 1063.5 | 1064.2 | 1076.2 |
| 31 | B$_2$ | 1125.0 | 1125.4 | 1129.7 |
| 32 | A$_1$ | 1159.6 | 1160.1 | 1164.5 |
| 33 | E    | 1464.1 | 1464.3 | 1467.3 |
| 34 | E    | 1464.1 | 1464.3 | 1467.3 |
| 35 | A$_1$ | 1501.8 | 1501.8 | 1503.8 |
| 36 | B$_2$ | 1559.6 | 1559.7 | 1563.4 |
| 37 | E    | 1604.5 | 1604.9 | 1607.2 |
| 38 | E    | 1604.5 | 1604.9 | 1607.2 |
| 39 | A$_1$ | 1618.5 | 1618.8 | 1619.6 |
| 40 | B$_2$ | 1653.0 | 1653.3 | 1654.4 |
| 41 | E    | 3337.3 | 3338.3 | 3381.0 |
| 42 | E    | 3337.3 | 3338.3 | 3381.0 |
| 43 | A$_1$ | 3363.8 | 3363.6 | 3342.9 |
| 44 | B$_2$ | 3364.7 | 3363.8 | 3386.1 |
| 45 | E    | 3468.5 | 3467.2 | 3517.1 |
| 46 | E    | 3468.5 | 3467.2 | 3517.1 |
| 47 | A$_1$ | 3469.4 | 3468.3 | 3488.4 |
| 48 | B$_2$ | 3502.2 | 3501.5 | 3525.4 |

Naphthalene

Table S20. Naphthalene's frequencies from Phonopy (harmonic Frequencies) and from the Frozen-Phonon Calculations (harmonic and anharmonic frequencies) at the Γ-point at PBE+D3M(BJ) level of theory (150 K-volume). All frequencies are given in cm$^{-1}$.

| Mode | Mulliken Symbol | Harmonic Frequency (Phonopy) | Harmonic Frequency (Frozen Phonon) | First Anharmonic Frequency (Frozen Phonon) |
|---|---|---|---|---|
| 4 | $B_u$ | 41.7 | 41.4 | 45.4 |
| 5 | $A_u$ | 49.4 | 49.5 | 49.4 |
| 6 | $B_u$ | 51.5 | 51.3 | 53.5 |
| 7 | $B_g$ | 67.4 | 66.9 | 67.0 |
| 8 | $A_u$ | 67.9 | 67.4 | 69.0 |
| 9 | $A_g$ | 72.3 | 72.2 | 74.1 |
| 10 | $B_u$ | 90.4 | 90.4 | 91.0 |
| 11 | $B_g$ | 101.9 | 102.5 | 105.4 |
| 12 | $A_g$ | 119.3 | 120.0 | 123.3 |
| 13 | $A_u$ | 173.7 | 173.6 | 177.1 |
| 14 | $A_g$ | 186.3 | 186.5 | 189.3 |
| 15 | $B_g$ | 203.1 | 203.2 | 205.0 |
| 16 | $B_u$ | 205.2 | 205.5 | 207.3 |
| 17 | $A_u$ | 355.9 | 356.0 | 356.7 |
| 18 | $A_u$ | 357.7 | 357.8 | 358.5 |
| 19 | $B_u$ | 383.3 | 383.5 | 384.5 |
| 20 | $B_u$ | 387.1 | 387.4 | 388.4 |
| 21 | $A_u$ | 462.2 | 462.2 | 463.3 |
| 22 | $B_g$ | 466.5 | 466.4 | 467.5 |
| 23 | $A_g$ | 467.2 | 467.7 | 470.8 |
| 24 | $A_g$ | 475.8 | 476.2 | 479.3 |
| 25 | $B_g$ | 504.1 | 504.2 | 504.3 |
| 26 | $A_u$ | 505.4 | 505.4 | 505.6 |
| 27 | $B_u$ | 509.1 | 509.2 | 509.3 |
| 28 | $B_g$ | 509.5 | 509.7 | 509.7 |
| 29 | $A_g$ | 613.4 | 613.6 | 613.8 |
| 30 | $B_g$ | 615.4 | 615.6 | 615.8 |
| 31 | $A_g$ | 621.0 | 621.2 | 621.3 |
| 32 | $B_u$ | 624.6 | 624.8 | 624.9 |
| 33 | $A_u$ | 710.4 | 710.7 | 721.5 |
| 34 | $A_u$ | 713.7 | 714.1 | 724.7 |
| 35 | $B_u$ | 762.8 | 762.9 | 763.3 |
| 36 | $A_g$ | 764.2 | 764.3 | 764.2 |
| 37 | $B_g$ | 768.4 | 768.7 | 769.2 |
| 38 | $B_g$ | 771.0 | 771.2 | 779.4 |
| 39 | $A_g$ | 771.9 | 772.2 | 772.4 |
| 40 | $A_g$ | 783.1 | 783.3 | 791.1 |
| 41 | $B_u$ | 790.5 | 790.5 | 790.6 |
| 42 | $B_g$ | 792.1 | 792.1 | 792.2 |
| 43 | $A_u$ | 835.0 | 835.6 | 843.8 |
| 44 | $A_u$ | 838.8 | 839.5 | 847.9 |
| 45 | $B_u$ | 875.2 | 875.6 | 879.4 |
| 46 | $A_u$ | 888.6 | 889.1 | 893.7 |
| 47 | $B_u$ | 924.0 | 924.1 | 924.2 |
| 48 | $A_g$ | 928.0 | 928.2 | 928.2 |
| 49 | $B_g$ | 937.6 | 937.8 | 942.8 |

| Mode | Mulliken Symbol | Harmonic Frequency (Phonopy) | Harmonic Frequency (Frozen Phonon) | First Anharmonic Frequency (Frozen Phonon) |
|---|---|---|---|---|
| 50 | $A_g$ | 941.9 | 942.2 | 947.0 |
| 51 | $B_g$ | 953.2 | 953.5 | 958.4 |
| 52 | $B_g$ | 953.4 | 953.7 | 959.2 |
| 53 | $A_g$ | 969.8 | 970.1 | 975.9 |
| 54 | $B_u$ | 975.5 | 975.8 | 980.9 |
| 55 | $A_u$ | 975.7 | 975.9 | 980.4 |
| 56 | $A_u$ | 978.4 | 978.6 | 983.5 |
| 57 | $B_u$ | 1015.2 | 1015.3 | 1016.2 |
| 58 | $A_g$ | 1015.5 | 1015.6 | 1016.5 |
| 59 | $B_g$ | 1020.6 | 1020.7 | 1021.2 |
| 60 | $B_u$ | 1024.2 | 1024.3 | 1025.4 |
| 61 | $A_u$ | 1114.6 | 1114.8 | 1115.8 |
| 62 | $A_g$ | 1114.7 | 1114.9 | 1115.8 |
| 63 | $B_g$ | 1134.6 | 1134.7 | 1137.5 |
| 64 | $A_u$ | 1137.4 | 1137.7 | 1139.5 |
| 65 | $B_u$ | 1139.5 | 1139.8 | 1142.3 |
| 66 | $A_g$ | 1141.5 | 1141.9 | 1146.0 |
| 67 | $B_g$ | 1145.8 | 1146.0 | 1148.3 |
| 68 | $A_u$ | 1148.2 | 1148.4 | 1151.9 |
| 69 | $A_g$ | 1215.5 | 1215.5 | 1216.5 |
| 70 | $B_u$ | 1216.4 | 1216.4 | 1217.4 |
| 71 | $B_g$ | 1234.3 | 1234.5 | 1235.7 |
| 72 | $A_u$ | 1236.4 | 1236.6 | 1237.8 |
| 73 | $B_u$ | 1254.7 | 1254.9 | 1255.4 |
| 74 | $B_g$ | 1258.8 | 1259.0 | 1259.6 |
| 75 | $A_g$ | 1373.4 | 1373.7 | 1374.6 |
| 76 | $A_u$ | 1373.5 | 1373.7 | 1374.7 |
| 77 | $B_u$ | 1377.6 | 1377.5 | 1379.9 |
| 78 | $B_u$ | 1379.5 | 1379.4 | 1381.8 |
| 79 | $A_u$ | 1387.9 | 1387.8 | 1390.4 |
| 80 | $A_u$ | 1388.6 | 1388.5 | 1390.3 |
| 81 | $B_u$ | 1441.8 | 1441.9 | 1442.6 |
| 82 | $B_g$ | 1445.5 | 1445.6 | 1446.4 |
| 83 | $A_g$ | 1447.4 | 1447.6 | 1448.4 |
| 84 | $B_g$ | 1448.7 | 1448.8 | 1449.6 |
| 85 | $B_g$ | 1504.3 | 1504.4 | 1505.3 |
| 86 | $A_g$ | 1506.5 | 1506.6 | 1507.5 |
| 87 | $A_g$ | 1564.2 | 1564.2 | 1565.1 |
| 88 | $B_u$ | 1566.7 | 1566.6 | 1567.6 |
| 89 | $A_u$ | 1587.9 | 1587.9 | 1589.2 |
| 90 | $A_g$ | 1588.5 | 1588.4 | 1589.7 |
| 91 | $B_g$ | 1619.1 | 1619.0 | 1619.6 |
| 92 | $A_u$ | 1619.5 | 1619.4 | 1620.0 |
| 93 | $B_u$ | 3088.2 | 3088.0 | 3101.1 |
| 94 | $B_g$ | 3088.9 | 3088.8 | 3100.7 |
| 95 | $A_g$ | 3089.5 | 3089.3 | 3101.7 |
| 96 | $A_u$ | 3090.6 | 3090.5 | 3102.6 |
| 97 | $A_g$ | 3099.2 | 3099.1 | 3108.1 |
| 98 | $B_u$ | 3099.4 | 3099.3 | 3108.1 |
| 99 | $B_g$ | 3100.1 | 3099.9 | 3109.0 |

| Mode | Mulliken Symbol | Harmonic Frequency (Phonopy) | Harmonic Frequency (Frozen Phonon) | First Anharmonic Frequency (Frozen Phonon) |
|---|---|---|---|---|
| 100 | $A_u$ | 3100.1 | 3100.0 | 3109.9 |
| 101 | $B_u$ | 3108.2 | 3108.0 | 3119.5 |
| 102 | $A_g$ | 3109.0 | 3108.8 | 3120.9 |
| 103 | $B_g$ | 3109.0 | 3108.8 | 3120.4 |
| 104 | $B_g$ | 3109.1 | 3108.9 | 3121.3 |
| 105 | $B_u$ | 3113.1 | 3113.0 | 3122.5 |
| 106 | $A_g$ | 3114.6 | 3114.6 | 3123.4 |
| 107 | $A_u$ | 3114.8 | 3114.8 | 3126.4 |
| 108 | $B_u$ | 3116.3 | 3116.2 | 3108.4 |